\documentclass{emulateapj}
\usepackage{natbib}
\usepackage{bm}
\usepackage{mathrsfs,amsmath}
\usepackage{graphicx}
\usepackage{epstopdf}
\usepackage[english]{babel}
\usepackage[colorlinks,linkcolor={blue},citecolor={blue},urlcolor={blue}]{hyperref}
\newcommand{\mpchi}{\,h^{-1}{\rm {Mpc}}}

\newcommand{\msun}{M_{\sun}}
\newcommand{\wrp}{w_{\rm p}}
\newcommand{\rp}{r_{\rm p}}
\newcommand{\wprp}{w_{\rm p}(r_{\rm p})}
\newcommand{\OII}{[O\,{\scriptsize II}]}

\shorttitle{Stellar mass functions of star-forming galaxies at $0.7<z<1.2$}
\shortauthors{H. Guo et al.}
\begin{document}
\title{Evolution of the Star-Forming Galaxies from $z=0.7$ to 1.2 with
  eBOSS Emission-line Galaxies }

\author{Hong Guo\altaffilmark{1}, Xiaohu Yang\altaffilmark{2,3}, Anand
  Raichoor\altaffilmark{4,5}, Zheng Zheng\altaffilmark{6}, Johan
  Comparat\altaffilmark{7}, V. Gonzalez-Perez\altaffilmark{8,9},
  Jean-Paul Kneib\altaffilmark{4,10}, Donald
  P. Schneider\altaffilmark{11,12}, Dmitry
  Bizyaev\altaffilmark{13,14}, Daniel Oravetz\altaffilmark{13}, Audrey
  Oravetz\altaffilmark{14}, Kaike Pan\altaffilmark{13}}

\altaffiltext{1}{Key Laboratory for Research in Galaxies and
  Cosmology, Shanghai Astronomical Observatory, Shanghai 200030,
  China; guohong@shao.ac.cn}

\altaffiltext{2}{Department of Astronomy, and Tsung-Dao Lee Institute,
  Shanghai Jiao Tong University, Shanghai 200240, China}

\altaffiltext{3}{IFSA Collaborative Innovation Center, and Shanghai
  Key Laboratory for Particle Physics and Cosmology, Shanghai Jiao
  Tong University, Shanghai 200240, China}

\altaffiltext{4}{Institute of Physics, Laboratory of Astrophysics,
  Ecole Polytechnique F\'ed\'erale de Lausanne (EPFL), Observatoire de
  Sauverny,CH-1290 Versoix, Switzerland}

\altaffiltext{5}{CEA, Centre de Saclay, IRFU/SPP, F-91191
  Gif-sur-Yvette, France}

\altaffiltext{6}{Department of Physics and Astronomy, University of
  Utah, UT 84112, USA}

\altaffiltext{7}{Max-Planck-Institut f\"ur extraterrestrische Physik
  (MPE), Giessenbachstrasse 1, D-85748 Garching bei M\"unchen,
  Germany}

\altaffiltext{8}{Institute for Computational Cosmology, Department of
  Physics, Durham University, South Road, Durham DH1 3LE, UK}

\altaffiltext{9}{Energy Lancaster, Lancaster University, Lancaster LA14YB, UK}

\altaffiltext{10}{Aix Marseille Universit\'e, CNRS, LAM (Laboratoire
  d'Astrophysique de Marseille), UMR 7326, F-13388 Marseille, France}

\altaffiltext{11}{Department of Astronomy and Astrophysics, The
  Pennsylvania State University, University Park, PA 16802, USA}

\altaffiltext{12}{Institute for Gravitation and the Cosmos, The
  Pennsylvania State University, University Park, PA 16802, USA}

\altaffiltext{13}{Apache Point Observatory and New Mexico State
  University, P.O. Box 59, Sunspot, NM, 88349-0059, USA}

\altaffiltext{14}{Sternberg Astronomical Institute, Moscow State
  University, Moscow, Russia}

\begin{abstract}
  We study the evolution of star-forming galaxies with
  $10^{10}\msun<M_*<10^{11.6}\msun$ over the redshift range of
  $0.7<z<1.2$ using the emission-line galaxies (ELGs) in the extended
  Baryon Oscillation Spectroscopic Survey (eBOSS). By applying the
  incomplete conditional stellar mass function (ICSMF) model proposed
  in \cite{Guo18}, we simultaneously constrain the sample
  completeness, the stellar--halo mass relation (SHMR) and the
  quenched galaxy fraction. We obtain the intrinsic stellar mass
  functions for star-forming galaxies in the redshift bins of
  $0.7<z<0.8$, $0.8<z<0.9$, $0.9<z<1.0$ and $1.0<z<1.2$, as well as
  the stellar mass function for all galaxies in the redshift bin of
  $0.7<z<0.8$. We find that the eBOSS ELG sample only
  selects about 1\%--10\% of the star-forming galaxy population at
  the different redshifts, with the lower redshift samples to be more
  complete. There is only weak evolution in the SHMR of the ELGs from
  $z=1.2$ to $z=0.7$, as well as the intrinsic galaxy stellar mass
  functions. Our best-fitting models
  show that the central ELGs at these redshifts live in halos of mass
  $M\sim10^{12}\msun$, while the satellite ELGs occupy slightly more
  massive halos of $M\sim10^{12.6}\msun$. The average satellite
  fraction of the observed ELGs varies from 13\% to 17\%, with the
  galaxy bias increasing from 1.1 to 1.4 from $z=0.7$ to 1.2.
\end{abstract}
	
	
\section{Introduction}

The next-generation large-scale galaxy redshift surveys will probe
much larger volumes into the deeper universe than the existing
surveys, e.g., the 2dF Galaxy Redshift Survey
\citep[2dFGRS;][]{Colless99} and the Sloan Digital Sky Survey
\citep[SDSS;][]{York00,Gunn06}. Efficient tracers of the large-scale
structure are necessary to probe the high-redshift universe. For
example, the Dark Energy Spectroscopic Instrument
\citep[DESI;][]{DESICollaboration16} is targeting the luminous red
galaxies (LRGs) up to $z=1.0$ and the star-forming galaxies with
strong nebular emission lines, a.k.a. emission-line galaxies
(ELGs), up to $z=1.7$. The Prime Focus Spectrograph
\citep[PFS;][]{Takada14} will target ELGs over a wide redshift range
of $0.8<z<2.4$ to constrain the cosmological parameters and study the
galaxy evolution. The 4-meter Multi-Object Spectroscopic Telescope
\citep[4MOST;][]{deJong16} also treats the ELGs as main targets for
cosmological probes at redshifts $0.7<z<1.2$. The Hobby-Eberly
Telescope Dark Energy Experiment \citep[HETDEX;][]{Hill08} will target
more than a million \OII\ emitting ELGs at $z<0.5$.

The \OII\ doublet emitters are of particular interest to these
high-redshift ELG surveys, as their strong emission lines at the
rest-frame wavelengths of $3727$ and $3729~\rm{\AA}$ will
make it easier to accurately measure redshifts beyond $z=1$, where the LRGs
are no longer efficient cosmological tracers \citep{Zhu15}. These
\OII\ emitters are also important tracers of the cosmic star formation
history \citep{Kewley04,Orsi14}, as the cosmic star formation rate
(SFR) peaks around $z\sim2$ \citep[see, e.g.,][]{Behroozi13,
  Yang13}. In addition, the broad stellar mass range probed by these
ELGs makes them good tracers of the galaxy stellar mass function
(SMF), especially the region around the knee of the SMF
\citep{Comparat17}.

Some of the physical and clustering properties of the ELGs have been
investigated in previous studies. For example, by combining the ELG
galaxy samples in the VIMOS VLT Deep Survey \citep[VVDS;][]{LeFevre13}
and the DEEP2 survey \citep{Newman13}, \cite{Comparat16a} found that
the characteristic luminosity of the \OII\ luminosity function
increases by a factor of 2.7 from $z=0.5$ to 1.3. \cite{Favole16}
constructed a sample of g-band-selected galaxies from the
Canada-France-Hawaii Telescope Legacy Survey
\citep[CFHT-LS;][]{Ilbert06}, which is expected to be dominated by
ELGs in the redshift range of $0.6<z<1$. They found that the typical
host halo mass of ELGs at $z\sim0.8$ is around $10^{12}h^{-1}\msun$
\citep[see also][]{Orsi14,Gonzalez-Perez18} and the satellite fraction
of the selected sample is about 22.5\%.

The number of observed high-redshift star-forming galaxies with
emission line measurements is growing quickly with recent optical and
near-infrared surveys \citep[see e.g.,][]{Comparat16, Okada16,
  Delubac17, Kaasinen17, Drinkwater18}. But the sample sizes of the
current ELG surveys are still not large enough to fully understand the
properties of the high-redshift ELGs. The SDSS-IV extended Baryon
Oscillation Spectroscopic Survey \citep[eBOSS;][]{Dawson16} has
recently finished its ELG survey program, and the final sample consists
of about 0.2 million \OII\ ELGs covering the redshift range of
$0.6<z<1.2$. Although the \OII\ emitters in eBOSS are mainly used as
the cosmological tracers \citep{Zhao16} and the galaxy spectra are
quite noisy, the large ELG sample provides an opportunity to better
understand the evolution of star-forming galaxies since $z=1.2$ and
the corresponding stellar-halo mass relations (SHMRs) \citep{Yang12,Behroozi13,Beutler13,Moster13,Lin16,Saito16}.

However, complicated target selections of ELGs in these high-redshift
cosmological surveys hinder the direct statistical studies of the
evolution of ELGs through cosmic time. It is hard to estimate the
sample completeness for these ELGs.  Using a semi-analytical model of
galaxy formation and evolution, \cite{Gonzalez-Perez18} found that the
sample completeness varies significantly in different surveys
and that the eBOSS ELG sample is highly incomplete at both the bright
and faint ends of the \OII\ luminosity function.

Several methods have been proposed to estimate the sample
completeness, e.g., by comparing the observed SMFs with those from
deeper imaging observations \citep{Leauthaud16,Saito16}, using galaxies
selected with relaxed color cuts \citep{Tinker17}, forward-modeling
the target selections with the analytical parametric maximum likelihood
method \citep{Montero-Dorta16}, or using the clustering redshift method by cross-correlating a spectroscopic sample with a parent photometric sample \citep{Bates18}. Recently, \cite{Guo18} (hereafter G18)
introduced a novel method of simultaneously constraining the galaxy
sample completeness and the SHMRs using the incomplete conditional
stellar mass function (ICSMF) model. It has the advantage of
estimating the sample completeness self-consistently using only the
observed galaxy samples, which is also in good agreement with the estimates from other methods \citep[see e.g.,][]{Leauthaud16,Bates18}. By applying the method to the SDSS-III Baryon
Oscillation Spectroscopic Survey \citep[BOSS;][]{Dawson13}, we found
that the intrinsic galaxy SMFs can be successfully derived from the
ICSMF model, which provides an efficient way of studying the evolution
of the galaxy SMF from these incomplete cosmological surveys.

In this paper, we will apply the ICSMF model to the final eBOSS ELG
sample in the redshift range of $0.7<z<1.2$ to constrain the sample
completeness, as well as to estimate their host halo masses. The
derived galaxy intrinsic SMFs for these star-forming galaxies allow us
to investigate the evolution of galaxy star formation histories.  The
eBOSS ELG sample used in this paper is defined as the \OII\ emitters
with at least one \OII\ emission line or the corresponding continuum
flux detected \citep[Eq.~1 of][]{Raichoor17}. The structure of this
paper is constructed as follows. In \S\ref{sec:data}, we describe the
galaxy samples and the simulation used in the modeling. We briefly
introduce our modeling method in \S\ref{sec:method} and present the
results for the eBOSS ELG in \S\ref{sec:results}. We discuss the
results in \S\ref{sec:discussion} and summarize in
\S\ref{sec:conclusion}.

Throughout this paper, we assume a spatially flat $\Lambda$CDM
cosmology, with $\Omega_{\rm m}=0.307$, $h=0.678$,
$\Omega_{\rm b}=0.048$ and $\sigma_8=0.823$, consistent with the
constraints from Planck \citep{PlanckCollaboration14} and with the
simulation used in our modeling (see \S\ref{sec:data}). For the galaxy
stellar mass estimates, we assume a universal \cite{Chabrier03}
initial mass function (IMF), the stellar population synthesis model of
\cite{Bruzual03} and the time-dependent dust attenuation model of
\cite{Charlot00}. All masses are in units of $\msun$.
	
\section{Data} \label{sec:data}

\subsection{eBOSS ELG Sample}

The eBOSS is one of three key surveys comprising SDSS-IV, aiming to
constrain the cosmological parameters at percent levels
\citep{Blanton17}. The ELGs are one of the four tracers of the underlying
matter density field in eBOSS. About 300 plates are dedicated to the
ELG observation, starting in 2016 \citep{Dawson16}. 

Though pilot surveys demonstrated that a target selection based on the SDSS imaging passed the eBOSS requirements \citep{Comparat16,Raichoor16,Delubac17}, the deep $grz$-band photometry of the Dark Energy Camera Legacy Survey (DECaLS\footnote{http://legacysurvey.org}) enables a more efficient target selection. Thus, the eBOSS ELG target selection \citep{Raichoor17} has been done on the DECaLS $grz$-band photometry. The \OII\ emitters are
selected with the following color and magnitude cuts. For the northern
galactic cap (NGC), the selection cuts are
\begin{small}
\begin{eqnarray}
21.825&<&g<22.9 \\
-0.068(r-z)+0.457&<&g-r<0.112(r-z)+0.773\\
0.637(g-r)+0.399&<&r-z<-0.555(g-r)+1.901,
\end{eqnarray}
\end{small}
while for the southern galactic cap (SGC), the cuts are changed to
\begin{small}
	\begin{eqnarray}
	21.825&<&g<22.825 \\
	-0.068(r-z)+0.457&<&g-r<0.112(r-z)+0.773\\
	0.218(g-r)+0.571&<&r-z<-0.555(g-r)+1.901.
	\end{eqnarray}
\end{small}
The different cuts for NGC and SGC are applied to account for the difference in imaging depth, the deeper SGC imaging permitting a selection closer to the low-redshift locus in the $grz$ diagram. The selection boxes in the $g-r$ and $r-z$ plane are adopted to maximize the fraction of $0.7<z<1.1$ \OII\ emitters. We refer the
readers to \cite{Raichoor17} for details (their Table~2 and Figure~4).

\begin{figure}
	\centering
	\includegraphics[width=0.47\textwidth]{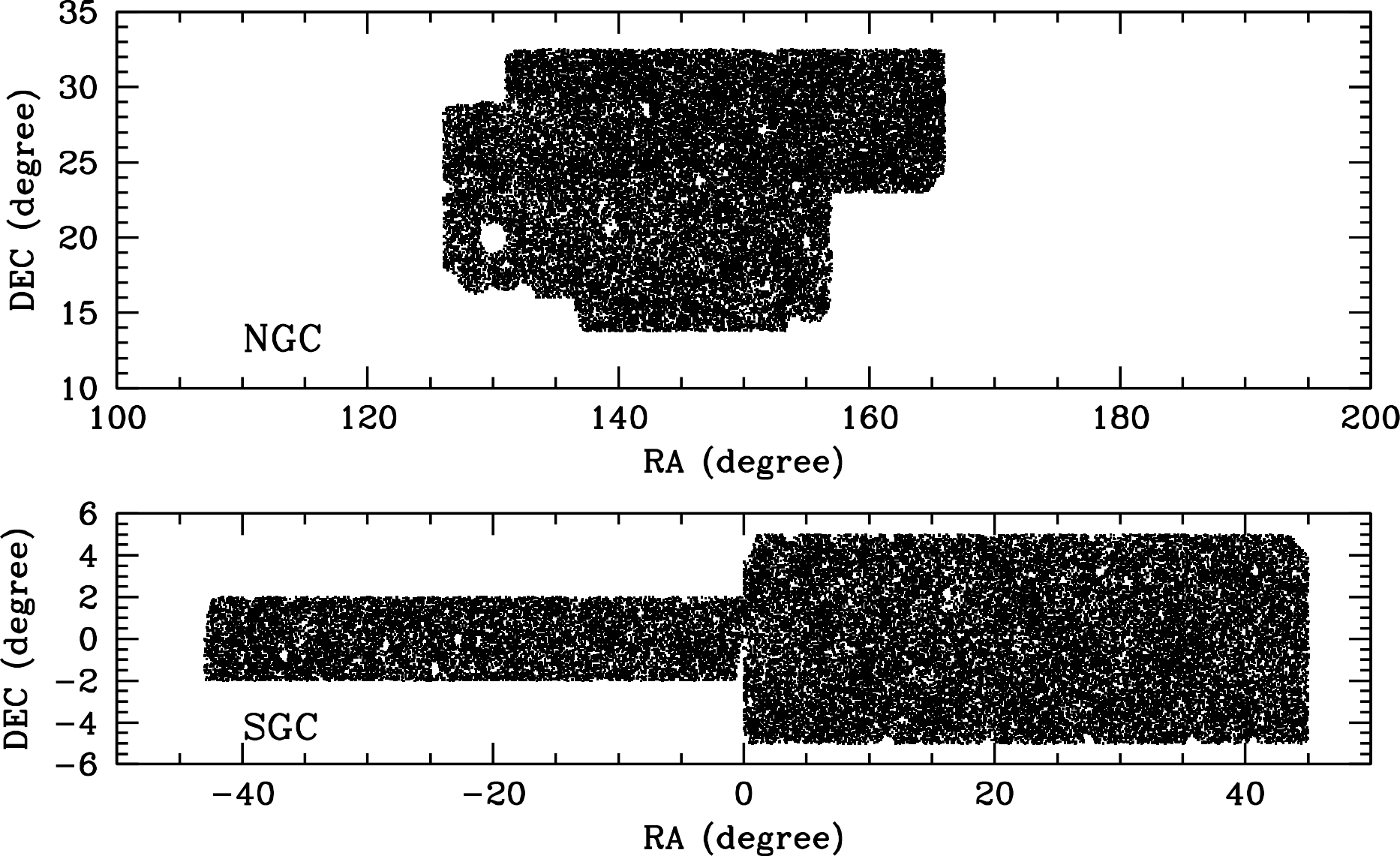}
	\caption{Angular distribution of the eBOSS ELG sample,
          separated into the northern (upper panel) and southern
          (lower panel) galactic caps. }
	\label{fig:radec}
\end{figure}

The observation of the eBOSS ELG program was completed in
2018 February. The 
final ELG sample consists of $\sim$222,000 galaxies with a reliable $z_{\rm spec}$ measurement; once all the angular masks are applied, the ELG footprint will cover a total area of $\sim830$ deg$^2$ (A. Raichoor et al. 2019, in preparation). We show
in Figure~\ref{fig:radec} the angular distribution of the ELG sample
for the NGC (upper panel) and SGC (lower panel), respectively.

\begin{figure}
	\centering
	\includegraphics[width=0.4\textwidth]{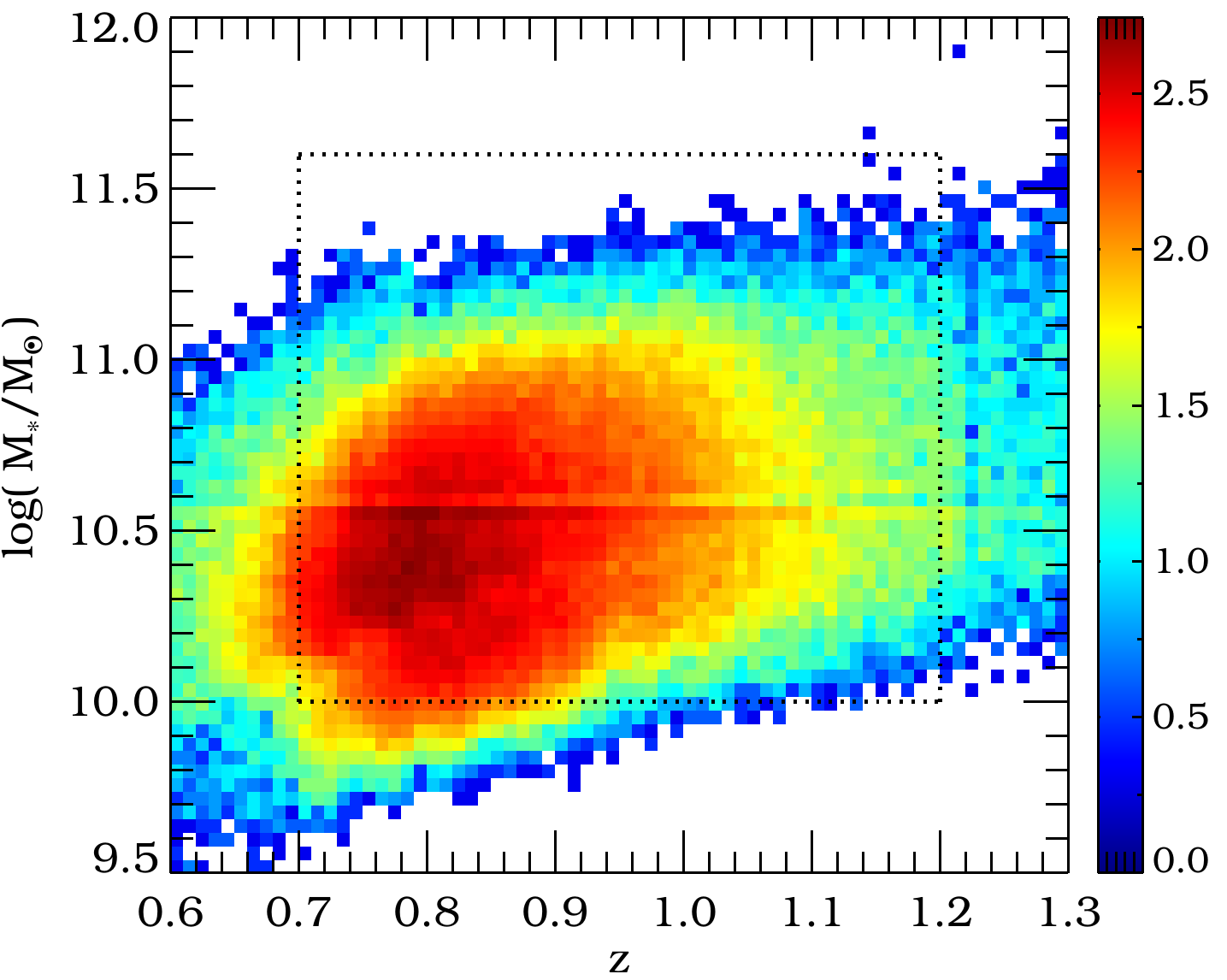}
	\caption{Distribution of the ELGs as a function of redshift
          and stellar mass. The dotted lines show the redshift and
          stellar mass cuts of the sample used in this paper. We are
          selecting galaxies in the redshift range of $0.7<z<1.2$ and
          stellar mass range of $10<\log(M_*/\msun)<11.6$. The color
          scales represent the logarithmic number counts at the
          redshift and stellar mass intervals.}
	\label{fig:mstar}
\end{figure}

The galaxy stellar mass is estimated for each object by performing the
spectral energy distribution (SED) fitting to the photometry of $grz$
bands in DECaLs and the $W1W2$ bands in Wide-field Infrared Survey
Explorer \citep[WISE;][]{Wright10} with the {\scriptsize FAST}
algorithm \citep{Kriek09}, assuming the \cite{Chabrier03} IMF, the
\cite{Bruzual03} SPS model, and the dust attenuation law of
\cite{Kriek13} (full details are provided in Section 6.3 of \citealt{Raichoor17}).  In order to match the stellar mass estimates of G18
with the same IMF and SPS model assumptions but with the dust
attenuation law of \cite{Charlot00}, we increase the stellar mass in
the ELG pipeline by 0.15~dex to take into account the difference in
the dust attenuation laws, as found by \cite{Perez-Gonzalez08}
\citep[see also][]{Rodriguez-Puebla17}.

The star formation rates (SFRs) of the ELGs are also available from the SED fittings of the {\scriptsize FAST} code. We check the reliability of the output SFR values by comparing them to the SFR estimates from those galaxies with high S/N H$\alpha$ and H$\beta$ line fluxes. After applying the extinction correction using the Balmer decrement following the dust attenuation law of \cite{Charlot00} and also the aperture correction of the finite fiber diameter, we can estimate the SFR for these galaxies using the H$\alpha$ luminosity following \cite{Kennicutt98}. We find that the SFR estimates from the {\scriptsize FAST} output and those from the H$\alpha$ luminosity are in good agreement with each other, with the average value of $\log(\rm{SFR}/\msun\,{\rm yr}^{-1})$ at $1<z<1.2$ being $1.63\pm0.43$. As shown in Figure~1 of \cite{Lapi17}, the previous UV observations of the star-forming galaxies would considerably underestimate the SFR function for galaxies with SFRs larger than $30\,\rm{\msun}\,{\rm yr}^{-1}$ as a result of the strong dust extinction. Therefore, the eBOSS ELG observation provides a valuable sample of dusty star-forming galaxies with moderate SFRs at $z\sim1$.

The stellar mass distribution of ELGs at different redshifts is
displayed in Figure~\ref{fig:mstar}. As for any other ELG sample for measuring the baryon acoustic oscillations, the eBOSS ELG target selection aims to select a homogeneous sample of strong \OII\ emitters with a given sky density within a given redshift range. As a consequence, the eBOSS ELG sample is not mass-complete. More massive ELGs are observed at higher redshifts by the g-band flux
limits. In this paper, we focus on the redshift range of $0.7<z<1.2$
and only use the ELGs with $10<\log(M_*/\msun)<11.6$. The redshift and
stellar mass selection cuts are shown as the dotted lines in
Figure~\ref{fig:mstar}.

In order to study the evolution of the SMF for the ELGs, we divide the
sample into four redshift bins from $z=0.7$ to 1.2, with a bin size of
$\Delta z=0.1$ for $z<1$ and a larger bin size of $1<z<1.2$ to achieve
enough signal-to-noise (S/N) for the clustering measurements. We
display the number of ELGs in each subsample, $N_{\rm tot}$, and the
corresponding number densities, $\bar{n}_{\rm g}$, in
Table~\ref{tab:data}.

\begin{deluxetable}{lcc}
	\centering
	\tablecaption{Samples of Different Redshift Bins \label{tab:data}} 
	\centering
	\tablehead{Redshift Range  & $N_{\rm tot}$ &$\bar{n}_{\rm g}/(h^3\, \rm{Mpc}^{-3})$}
	\startdata
	$0.7<z<0.8$  & 61197 & $3.61\times10^{-4}$\\			
	$0.8<z<0.9$  & 71172 & $3.66\times10^{-4}$\\	
	$0.9<z<1.0$  & 37025 & $1.71\times10^{-4}$\\	
	$1.0<z<1.2$  & 24232 & $0.49\times10^{-4}$	                 			
	\enddata
	\tablecomments{The total number of galaxies and the average
          galaxy number densities of the eBOSS ELG samples at
          different redshifts are displayed.}
\end{deluxetable}

\subsection{Dark Matter Simulation}

To apply the ICSMF model, we directly use the dark matter halo
catalogs from the Multidark Planck simulation
\citep[MDPL\footnote{https://www.cosmosim.org/cms/simulations/mdpl2/};][]{Klypin16},
with the cosmological parameters of $\Omega_{\rm m}=0.307$,
$\Omega_{\rm b}=0.048$, $h=0.678$, $n_{\rm s}=0.96$ and
$\sigma_8=0.823$. The simulation has a box size of
$1\,h^{-1}{\rm{Gpc}}$ and a mass resolution of
$1.5\times10^{9}h^{-1}M_\odot$. The simulation resolution is high
enough to resolve the host halos for the ELGs, which is around
$10^{12}\msun$ \citep{Favole16}. The dark matter halos and subhalos in
the simulation are identified with the {\scriptsize ROCKSTAR}
phase-space halo finder \citep{Behroozi13b}. We use four different
redshift outputs of $z=0.740$, 0.859, 0.944 and 1.077 from MDPL,
roughly corresponding to the median redshifts of the four ELG
subsamples.

\section{Method} \label{sec:method}

In this section, we briefly introduce the main ingredients of the
ICSMF model of G18, which is based on the traditional conditional
stellar mass function (CSMF) framework \citep[see
e.g.,][]{Yang12,vandenBosch13}. By incorporating the stellar mass
completeness, the ICSMF model is able to use the observed (incomplete)
galaxy SMF and the two-point correlation function (2PCF) to
simultaneously constrain the sample completeness and the galaxy
SHMR. We refer the readers to G18 for more details.

\subsection{ICSMF Model Ingredients}\label{subsec:model}

The three key ingredients of the ICSMF model are the CSMF, the SHMR
and the stellar mass incompleteness. As in \cite{Yang12} and G18, we
assume a log-normal distribution for the central galaxy CSMF, i.e.,
the average number of central galaxies with stellar mass $M_*$ in host
halos of given mass $M$,
\begin{equation}
  \Phi_{\rm c}^{\rm sf}(M_*|M)=\frac{1}{\sqrt{2\pi}\sigma_{\rm *}}\exp
  \left[-\frac{(\log M_*- \log\langle M_*|M\rangle)^2} {2 \sigma_{\rm *}^2}\right] \label{eq:csmf}
\end{equation}
where $\sigma_{\rm *}$ characterizes the scatter of galaxy stellar
mass at a given halo mass and the function $\langle M_*|M\rangle$ is
the average central galaxy stellar mass in halos of mass $M$, i.e. the
SHMR. Following \cite{Yang12}, we assume a constant scatter
$\sigma_{\rm *}$ of $\max(0.173,0.2z)$ at a given redshift $z$ and the
functional form for $\langle M_*|M\rangle$ is assumed to be a broken
power law, \citep{Yang09,Wang10}
\begin{equation}
 \langle M_*|M\rangle = M_{*,0}\frac{(M/M_1)^{
  		\alpha+\beta}}{(1+M/M_1)^{\beta}} \label{eq:smhm}
\end{equation}
where $M_{*,0}$, $M_1$, $\alpha$, and $\beta$ are the four model
parameters. The values of $\alpha+\beta$ and $\alpha$ represent the
slopes of the low- and high-mass ends of the SHMR, respectively.

In G18, the stellar mass completeness is decomposed into the
contributions from the central and satellite galaxies. After trying
the model as in G18, we find that for the eBOSS ELG samples, the
separate contributions of the central and satellite completeness
functions are not well constrained, and they have similar completeness
functions. Therefore, in this paper, we only assume an overall
stellar mass completeness function, $c(M_*)$, for all galaxies, as
follows \citep{Leauthaud16},
\begin{equation}
  c(M_*)=\frac{f_{\rm c}}{2}\left[1+{\rm erf}\left(\frac{\log M_* 
        - \log M_{\rm *,c}}{\sigma_{\rm c}}\right)\right] \label{eq:cmstar}
\end{equation}
where $\rm{erf}$ is the error function and the three free parameters
are $f_{\rm c}$, $M_{\rm *,c}$, and $\sigma_{\rm c}$.

In G18, the ICSMF model was applied to the galaxies in the SDSS-III
Baryon Oscillation Spectroscopic Survey \citep[BOSS;][]{Dawson13} at
$0.1<z<0.8$ for $10^{11}\msun<M_\ast<10^{12}\msun$, where the
red/quiescent galaxies dominate the galaxy SMF. However, for the eBOSS ELG
sample, the galaxy stellar mass spans two orders of magnitudes from
$10^{10}\msun$ to $10^{12}\msun$, where the star-forming galaxies are
not always dominating the entire galaxy population. In order to
properly model the eBOSS ELGs that are mostly star-forming galaxies
\citep{Zhu15,Comparat16}, we need to quantify the number of
star-forming galaxies in a given halo, which requires the measurements
for the quiescent galaxies. Since the fraction of quiescent galaxies
(i.e., quenched fraction) does not have a strong evolution in
$0.7<z<1.2$ \citep[see e.g.,][]{Moustakas13,Muzzin13,Tomczak14}, it is
possible to constrain the quenched fraction using both the eBOSS ELGs
and the BOSS LRGs in the overlap redshift range of $0.7<z<0.8$.

Here, we assume that the quenched fraction $f_{\rm q}$ is a function
of the host halo mass \citep{Tinker13,Rodriguez-Puebla15,Zu16}, which provides more
flexiblity in the modeling as the star-forming and quenched galaxies
could have different SHMRs. We adopt a similar functional form as in
\cite{Peng12} and \cite{Rodriguez-Puebla15} for quenched ($f_{\rm q}$) and star-forming galaxies
($f_{\rm sf}$) as follows,
\begin{eqnarray}
f_{\rm q}(M)&=&\frac{1}{1+M/M_{\rm q}}\label{eq:fsf},\\
f_{\rm sf}(M)&=&1-f_{\rm q}(M)
\end{eqnarray}
The free parameter $M_{\rm q}$ characterizes the mass scale where half
of the halos at a given mass $M$ contain quenched central galaxies. We
note that with the incorporation of the quenched fraction, $c(M_*)$ is
just the completeness function relative to the star-forming galaxy
population, but not to the whole population. We will adopt the same value
of $M_{\rm q}$ constrained by the joint modeling of the eBOSS ELGs and
BOSS LRGs in $0.7<z<0.8$ for all higher redshift samples, as will be detailed in
\S\ref{sec:fitting}.

\subsection{Modeling the 2PCF Measurements}

With the above four ingredients, we are able to predict the incomplete
halo (subhalo) occupation functions for the central (satellite) ELGs
in the stellar mass range of $M_{*,1}<M_*<M_{*,2}$, as
\begin{eqnarray}
    \langle N_{\rm c}(M)\rangle&=&\int_{M_{*,1}}^{M_{*,2}} \Phi_{\rm c}^{\rm sf}(M_*|M)c(M_*) f_{\rm sf}(M)dM_* \label{eq:ncen_int}\\
    \langle N_{\rm s}(M_{\rm acc})\rangle&=&\int_{M_{*,1}}^{M_{*,2}}
      \Phi_{\rm c}^{\rm sf}(M_*|M_{\rm acc}) c(M_*) f_{\rm sf}(M_{\rm acc})dM_* \nonumber\\
      \label{eq:nsat_int}
\end{eqnarray}
where the subscripts `c' and `s' are for the central and satellite
galaxies, respectively, and $M_{\rm acc}$ is the subhalo mass at the
last accretion epoch. We have assumed that the satellite galaxies have
the same CSMF as the centrals when they are distinct halos at the last
accretion epoch. The possible evolution of
$\Phi_{\rm c}^{\rm sf}(M_*|M)$ and the satellite galaxy stellar mass after
accretion has been ignored \citep[see][for a more sophisticated
model]{Yang12}.

To compare with the traditional halo occupation distribution (HOD)
\citep[see e.g.,][]{Zheng05,Zheng07,Zehavi11}, we can estimate the
occupation function of the satellite galaxies in the host halo,
$\langle N_{\rm s}(M)\rangle$, as follows:
\begin{eqnarray}
    \langle N_{\rm s}(M)\rangle&=&\int_{M_{*,1}}^{M_{*,2}} \Phi_{\rm s}^{\rm sf}(M_*|M)c(M_*) f_{\rm sf}(M)dM_*\label{eq:nsat}\\
\Phi_{\rm s}^{\rm sf}(M_*|M)&=&\int dM_{\rm acc} \Phi_{\rm c}^{\rm sf}(M_*|M_{
	\rm acc})n_{\rm s}(M_{\rm acc}|M)\,, \label{eq:csmfs}
\end{eqnarray}
where $\Phi_{\rm s}^{\rm sf}(M_*|M)$ is the CSMF for the satellite
galaxies and $n_{\rm s}(M_{\rm acc}|M)$ is the subhalo mass function
in host halos of mass $M$.

With the central and satellite HODs, the galaxy 2PCFs and the observed
ELG SMFs can be predicted with the MDPL halo and subhalo catalogs in
order to constrain the model parameters. We apply the efficient
simulation-based method of \cite{Zheng16} to compute the 3D galaxy
2PCF, $\xi(r_{\rm p},r_{\rm\pi})$, where $r_{\rm\pi}$ and $r_{\rm p}$
are the separations of galaxy pairs along and perpendicular to the
line of sight (LOS).

In more detail, the 3D galaxy 2PCF $\xi({\mathbf r})$ is measured as
\begin{eqnarray}
&&\xi({\mathbf r})=\sum_{i,j}\frac{n_{{\rm h},i}n_{{\rm h},j}}{\bar{n}_{\rm g}^2}
\langle N_{\rm c}(M_i)\rangle\langle N_{\rm c}(M_j)\rangle\xi_{\rm hh}({\mathbf r};M_i,M_j)\nonumber \\
&+&\sum_{i,j}2\frac{n_{{\rm h},i}n_{{\rm s},j}}{\bar{n}_{\rm g}^2}
\langle N_{\rm c}(M_i)\rangle\langle N_{\rm s}(M_{{\rm acc},j}) \rangle\xi_{\rm hs}({\mathbf r};M_i,M_{{\rm acc},j})\nonumber \\
&+&\sum_{i,j}\frac{n_{{\rm s},i}n_{{\rm s},j}}{\bar{n}_{\rm g}^2}
\langle N_{\rm s}(M_{{\rm acc},i})\rangle\langle N_{\rm s}(M_{{\rm
		acc},j})\rangle \xi_{\rm ss}({\mathbf r};M_{{\rm acc},i},M_{{\rm acc},j})\nonumber\\ \label{eq:xicc_sim}
\end{eqnarray}
where $n_{\rm h}(M)$ and $n_{\rm s}(M_{\rm acc})$ are the halo and subhalo mass functions,
respectively, with $i$ and $j$ for different halo mass bins. The galaxy number density $\bar{n}_{\rm g}$ is computed as
\begin{equation}
\bar{n}_{\rm g}=\sum_i\left[\langle N_{\rm c}(M_i)\rangle n_{\rm
	h}(M_i)+ \langle N_{\rm s}(M_{{\rm acc},i})\rangle n_{\rm s}(M_{{\rm acc},i})\right]. \label{eq:nc}
\end{equation}  
The 3D 2PCFs $\xi_{\rm hh}({\mathbf r};M_i,M_j)$,
$\xi_{\rm hs}({\mathbf r};M_i,M_{{\rm acc},j})$, and
$\xi_{\rm ss}({\mathbf r};M_{{\rm acc},i},M_{{\rm acc},j})$ are the
tabulated 2PCFs of the halo--halo, halo--subhalo, and subhalo--subhalo
pairs, measured directly in the simulation.

To reduce the effect of redshift-space distortion (RSD), we focus on the measurements of the
projected 2PCF $w_{\rm p}(r_{\rm p})$ \citep{Davis83}, defined as
\begin{equation}
w_{\rm p}(r_{\rm p})=2\int_{0}^{r_{\rm \pi,max}} \xi(r_{\rm p},r_{\rm\pi})dr_{\rm\pi}, \label{eq:wp}
\end{equation}
where $r_{\rm \pi,max}$ is the maximum LOS distance to achieve the best S/N.

\begin{figure*}
	\centering
	\includegraphics[width=0.8\textwidth]{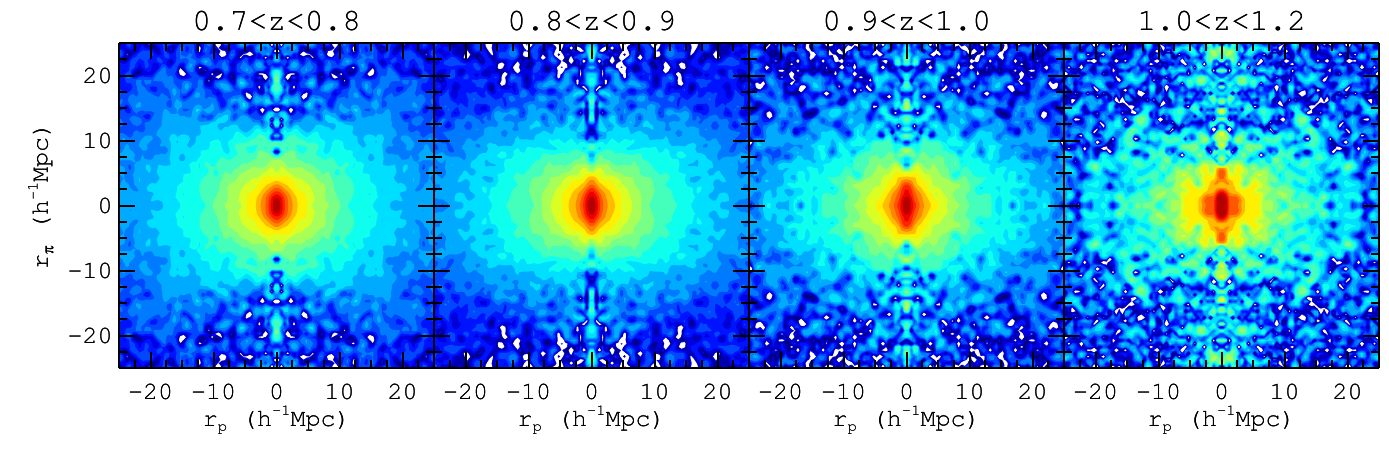}
	\caption{Measurements of the 3D 2PCF $\xi(\rp,r_{\rm\pi})$ for
          the eBOSS ELGs at different redshifts. Despite the noisy
          signals on large scales, the small-scale Fingers-of-God and
          large-scale Kaiser squashing effects are clearly shown. Most
          of the clustering signals are contained in the LOS distance
          of $20\mpchi$, which is set as the maximum integration
          distance of the projected 2PCF measurements.}
	\label{fig:xiplinear}
\end{figure*}

\subsection{Modeling Observed Galaxy SMFs}

By decomposing it into the central and satellite contributions, the
observed (incomplete) galaxy SMF of the ELGs can be predicted to be
\begin{eqnarray}
\Phi_{\rm sf}(M_*)&=&\Phi_{\rm sf,c}(M_*)+\Phi_{\rm sf,s}(M_*)\\
\Phi_{\rm sf,c}(M_*)&=&\int dM \Phi_{\rm c}^{\rm sf}(M_*|M) c(M_*) f_{\rm sf}(M) n_{\rm h}(M) \\
\Phi_{\rm sf,s}(M_*)&=&\int dM_{\rm acc} \Phi_{\rm c}^{\rm sf}(M_*|M_{\rm acc}) c(M_*)\times\nonumber\\
&&\quad\quad f_{\rm sf}(M_{\rm acc})n_{\rm s}(M_{\rm acc}) \label{eq:phisat}
\end{eqnarray}

In summary, we have four free parameters ($M_{*,0}$, $M_1$, $\alpha$,
and $\beta$) for the SHMR (Eq.~\ref{eq:smhm}),
another three parameters ($f_{\rm c}$, $M_{\rm *,c}$,
$\sigma_{\rm c}$) for the incompleteness component, and one parameter
($M_{\rm q}$) for the quenched fraction. The predictions of
$w_{\rm p}(r_{\rm p})$ and $\Phi_{\rm sf}(M_*)$ can be compared with
those measured in the observed galaxy samples to obtain the
best-fitting model parameters.

\subsection{Observational Measurements}
\label{sec:obs_meas}

The 3D galaxy 2PCF $\xi(r_{\rm p},r_{\rm\pi})$ and the projected 2PCF
$\wprp$ for the ELGs are measured with the Landy--Szalay estimator
\citep{Landy93}. We show in Figure~\ref{fig:xiplinear} the
measurements of $\xi(\rp,r_{\rm\pi})$ for samples at the different
redshifts. Although the measurements of $\xi(r_{\rm p},r_{\rm\pi})$
are noisy on most scales, there are apparent Fingers-of-God
\citep{Jackson72} and Kaiser squashing effects \citep{Kaiser87} on
small and large scales, respectively. Similar to the measurements of
\cite{Guo13} for BOSS galaxies at $z\sim0.55$, most of the clustering
signals for the ELGs are located within an LOS distance of
$20\mpchi$. Therefore, in order to achieve the best S/N ratio
especially for the high-redshift ELG samples, we only integrate
$\xi(r_{\rm p},r_{\rm\pi})$ to $r_{\pi,{\rm max}}=20\mpchi$ for the
measurements of $\wprp$. The residual RSD effect is taken into account
in the theoretical model predictions with the same LOS distance in
Eq.~\ref{eq:wp}.

We choose logarithmic $r_{\rm p}$ bins with a width
$\Delta\log r_{\rm p}=0.2$ from $1$ to $63.1\mpchi$, and linear
$r_{\rm\pi}$ bins of width $\Delta r_{\rm\pi}=2\mpchi$ from 0 to
20$\mpchi$. We measure the projected 2PCFs for the three stellar mass
bins from $M_*=10^{10}\msun$ to $10^{11.5}\msun$ with a bin size of
$\Delta\log M_*=0.5$ at different redshifts. The observed SMF
$\Phi_{\rm sf}(M_*)$ is measured in the stellar mass range of
$10^{10}\msun<M_*<10^{11.6}\msun$ with a logarithmic width of
$\Delta\log M_*=0.2$.

\subsection{ICSMF Model Constraints}
\label{sec:mod_cons}

We estimate the error covariance matrices for $w_{\rm p}(r_{\rm p})$
and $\Phi_{\rm sf}(M_*)$ using the jackknife resampling technique
with 100 subsamples as in G18. The cross-covariance between the
$w_{\rm p}(r_{\rm p})$ measurements for the different stellar mass
bins are also taken into account in the full covariance matrix. We
only use the diagonal elements of the covariance matrix of
$\Phi_{\rm sf}$, as the uncertainties from the systematic effects of the
stellar mass measurements are hard to estimate \citep{Mitchell13}. The
contribution of Poisson noise to the observed ELG SMF is added in
quadrature to $\sigma_{\Phi_{\rm sf}}$.

We apply a Markov Chain Monte Carlo (MCMC) method to fully explore the
model parameter space. The probability likelihood surface is
determined by $\chi^2$ as follows:
\begin{eqnarray}
\chi^2&=&\chi^2_{\wrp}+\frac{(\Phi_{\rm sf}-\Phi_{\rm sf}^*)^2}{\sigma_{\Phi_{\rm sf}}^2}\\
\chi^2_{\wrp}&=& (\mathbf{\wrp}-\mathbf{\wrp^*})^T
                 \mathbf{C}_{\wrp}^{-1} (\mathbf{\wrp}-\mathbf{\wrp^*}),
\end{eqnarray}
where $\mathbf{C}_{\wrp}$ is the full error covariance matrix of
$\wprp$. The quantity with (without) a superscript `$*$' is the one
from the data (model).

\subsection{ICSMF Model Predictions} \label{subsec:predictions}

With the best-fitting model constraints, we can also infer other
properties of the ELG samples, e.g., the galaxy bias, the satellite
fraction and the intrinsic galaxy SMFs. The galaxy bias for the
stellar mass range of $M_{*,1}<M_*<M_{*,2}$ can be directly estimated
as \citep{vandenBosch13},
\begin{eqnarray}
b_{\rm g}(M_*)&=&\frac{1}{\bar{n}_{\rm g}}\int dM \langle N(M)\rangle n_{\rm
	h}(M)b_{\rm h}(M) \label{eq:bias} \\
\langle N(M)\rangle &=& \langle N_{\rm c}(M)\rangle+\langle N_{\rm s}(M)\rangle.
\end{eqnarray}  
We adopt the halo bias fitting function of \cite{Tinker10} for
$b_{\rm h}(M)$ \citep[see also][]{Comparat17a} and
$\langle N(M)\rangle$ is the average halo occupation number for
galaxies in this stellar mass bin. The average galaxy bias
$\langle b_{\rm g}\rangle$ of each galaxy sample can be obtained by
using the halo occupation numbers for the whole stellar mass range in
Equations.~\ref{eq:ncen_int} and ~\ref{eq:nsat_int}.

The satellite galaxy fraction, $f_{\rm sat}(M_*)$, in the observed ELG
sample can be estimated as,
\begin{equation}
f_{\rm sat}(M_*)=\Phi_{\rm sf,s}(M_*)/\Phi_{\rm sf}(M_*),
\end{equation}
The average satellite fraction of the whole galaxy sample at a given
redshift interval is
\begin{equation}
\langle f_{\rm sat}\rangle=\frac{\int dM_*\Phi_{\rm sf,s}(M_*)}{\int dM_*\Phi_{\rm sf}(M_*)}\, .
\end{equation}
Most importantly, we can infer the intrinsic galaxy SMF for the
star-forming population, $\tilde{\Phi}_{\rm sf}(M_*)$, to be
\begin{eqnarray}
\tilde{\Phi}_{\rm sf}(M_*)=\int dM \Phi_{\rm c}^{\rm sf}(M_*|M) f_{\rm sf}(M) n_{\rm h}(M)\nonumber\\
+\int dM_{\rm acc} \Phi_{\rm c}^{\rm sf}(M_*|M_{\rm acc}) f_{\rm sf}(M_{\rm acc}) n_{\rm s}(M_{\rm acc})\,.
\end{eqnarray}
We can further predict the intrinsic galaxy SMF for the quenched
galaxies, $\tilde{\Phi}_{\rm q}(M_*)$, and that of the total
population, $\tilde{\Phi}(M_*)$, as,
\begin{eqnarray}
&&\tilde{\Phi}_{\rm q}(M_*)=\int dM \Phi_{\rm c}^{\rm q}(M_*|M) f_{\rm q}(M) n_{\rm h}(M)\nonumber\\
&&\quad+\int dM_{\rm acc} \Phi_{\rm c}^{\rm q}(M_*|M_{\rm acc}) f_{\rm q}(M_{\rm acc}) n_{\rm s}(M_{\rm acc})\\
&&\tilde{\Phi}(M_*)=\tilde{\Phi}_{\rm sf}(M_*)+\tilde{\Phi}_{\rm q}(M_*),
\end{eqnarray}
where we have assumed different conditional stellar mass functions,
$\Phi_{\rm c}^{\rm sf}(M_*|M) $ and $\Phi_{\rm c}^{\rm q}(M_*|M) $,
for tje star-forming and quenched galaxies, respectively.

In principle, the quenched and star-forming galaxies can have
different SHMRs. However, the level of differences is still under
debate in the literature, even for the well-measured galaxy populations
from the SDSS main galaxy sample at $z\sim0$ \citep[see][for a review
and references therein, especially their Figure~10]{Wechsler18}. It is
still not quite clear whether the SHMRs of the star-forming and
quenched galaxies are significantly different from each other at high
redshifts \citep[see e.g.,][for an analysis using samples with
photometric redshifts]{Tinker13}. In this study, with both the eBOSS
ELGs and BOSS LRGs at the same redshift range of $0.7<z<0.8$, we are
able to discriminate the possible differences between the CSMFs of the
two populations.

\section{Results}\label{sec:results}

\subsection{Fitting the Observables via two steps}\label{sec:fitting}

\begin{figure*}
	\centering
	\includegraphics[width=0.8\textwidth]{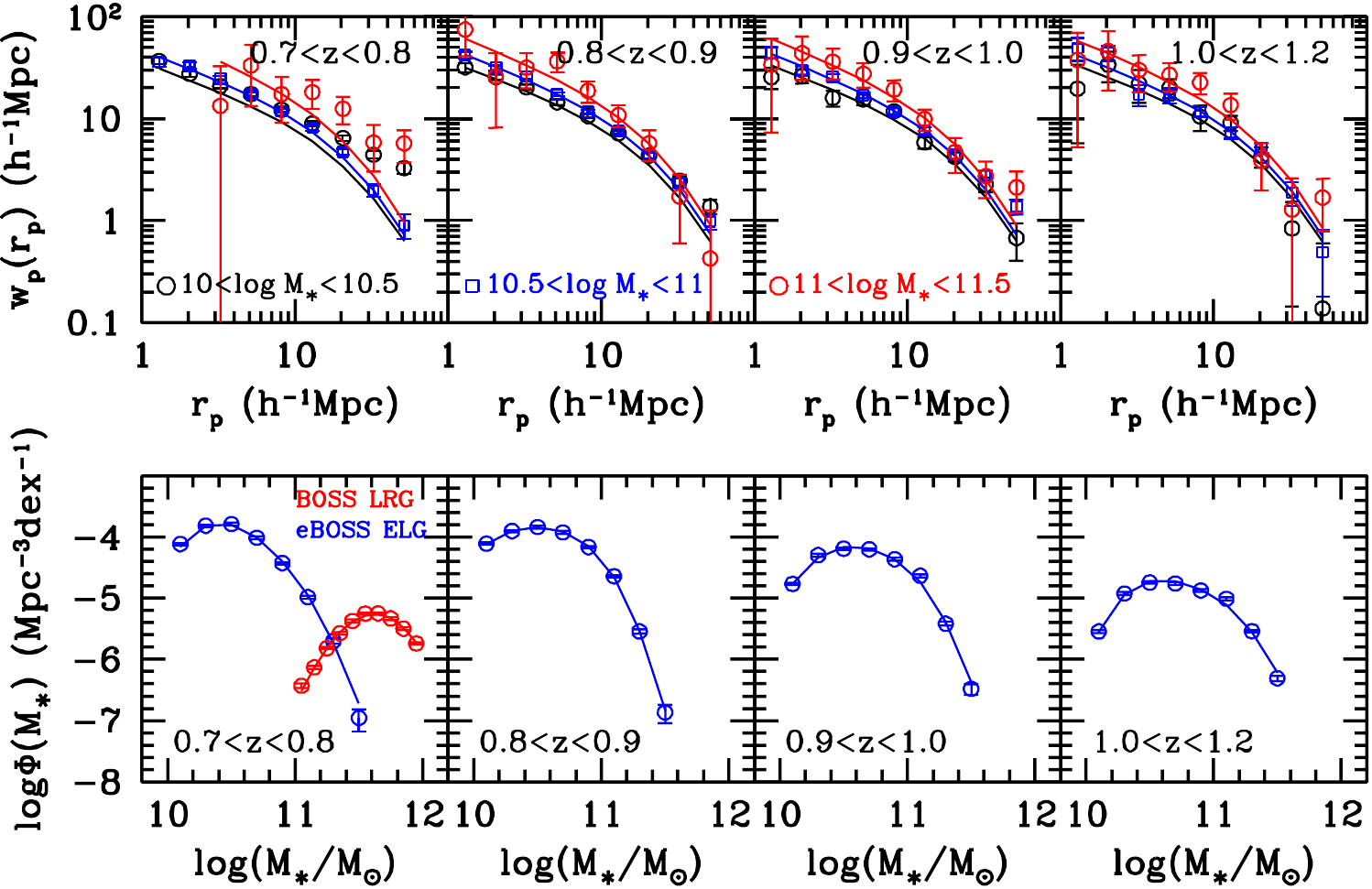}
	\caption{Projected 2PCF measurements (upper panels) and the
          observed SMFs (lower panels) for the eBOSS ELGs at
          $0.7<z<1.2$. In the upper panels, the symbols with different
          colors represent the $\wprp$ for the different stellar mass bins as
          labeled, and the corresponding solid lines are the
          best-fitting models. In the lower panels, the observed SMF
          measurements and the best-fitting models are shown as the
          symbols and lines, respectively. For the redshift bin of
          $0.9<z<1$, only nine data points are shown, as there are no
          observed ELGs in the largest stellar mass bin. Our
          best-fitting models are in very good agreement with the
          observed clustering and SMF measurements.}
	\label{fig:wp}
\end{figure*}

\begin{figure}
	\centering
	\includegraphics[width=0.35\textwidth]{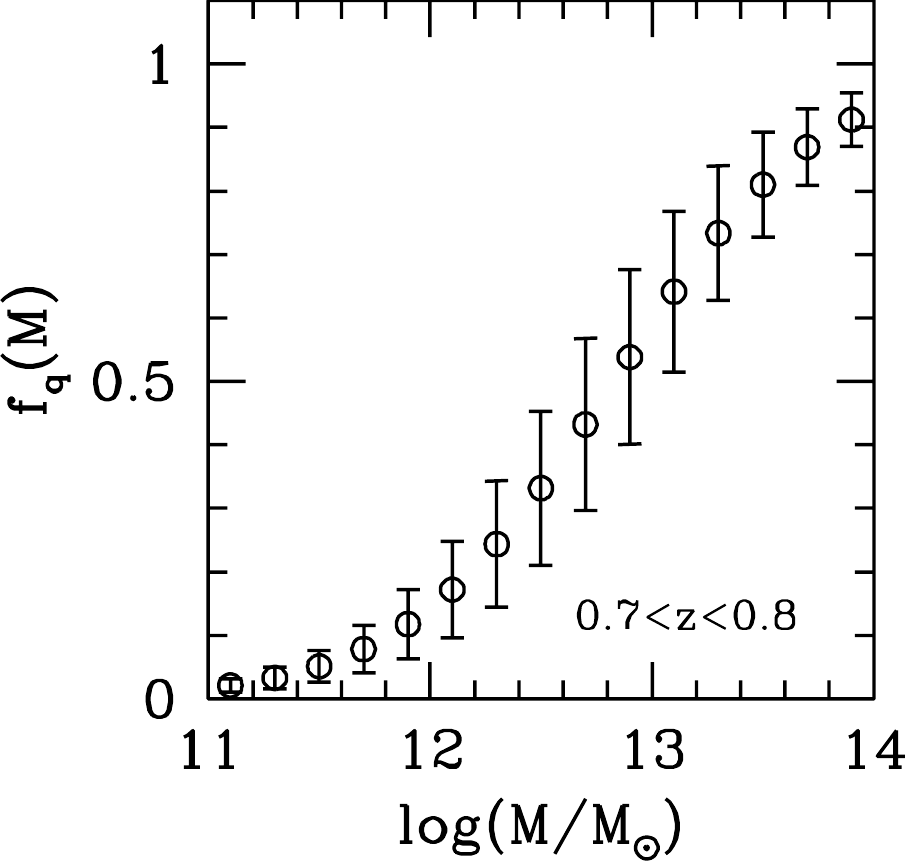}
	\caption{Best-fitting quenched fraction as a function of halo
          mass $M$ (open circles with errors). }
	\label{fig:fred}
\end{figure}

We show in Figure~\ref{fig:wp} the projected 2PCF measurements (upper
panels) and the observed SMFs (lower panels) for the eBOSS ELGs at
$0.7<z<1.2$. In the upper panels, the symbols of different colors are
for the ELGs in the different stellar mass bins.  The clustering
amplitudes of $\wprp$ for the two stellar bins of
$\log (M_*/\msun)<11$ almost overlap with each other,
reflecting the flattening trend of the galaxy bias at the lower mass
end, similar to the situation in the SDSS main galaxy sample
\citep[see e.g.,][]{Li06,Zehavi11}. The observed ELG SMFs (blue
circles) are shown in the lower panel of Figure~\ref{fig:wp}.

Before we fit to the data in all redshift bins, as we pointed out in
\S\ref{subsec:model}, the quenched fraction $f_{\rm q}(M)$ can be
properly constrained by jointly modeling the eBOSS ELG and BOSS LRG
samples. As these data are available only in the redshift interval of
$0.7<z<0.8$, we first focus on this redshift bin to make our model
constraints.

\begin{deluxetable*}{lrrrr}
	\centering	
	\tabletypesize{\scriptsize}
	\tablecaption{Best-fitting Model Parameters \label{tab:model}} 
	\tablehead{Parameter & $0.7<z<0.8$ & $0.8<z<0.9$ & $0.9<z<1.0$ & $1.0<z<1.2$}
	
	\startdata
	$\chi^2/{\rm dof}$ & $118.22/48$ & $25.52/28$ & $48.43/28$ & $46.16/28$ \\
	
	$\log (M_{*,0}/\msun)$ & $10.231^{+0.233}_{-0.087}$ & $11.006^{+0.046}_{-0.310}$ & $11.090^{+0.014}_{-0.122}$ & $11.199^{+0.016}_{-0.113}$    
	\\
	
	$\log (M_1/\msun)$ & $10.893^{+0.127}_{-0.168}$ & $11.342^{+0.008}_{-0.202}$ & $11.408^{+0.046}_{-0.080}$ & $11.454^{+0.015}_{-0.120}$ 
	\\
	
	$\alpha$ & $ 0.346^{+0.003}_{-0.120}$ & $ 0.020^{+0.144}_{-0.020}$ & $ 0.007^{+0.066}_{-0.007}$ & $ 0.007^{+0.067}_{-0.007}$ 
	\\
	
	$\beta$ & $ 7.729^{+0.489}_{-0.024}$ & $ 8.094^{+0.596}_{-0.726}$ & $ 8.574^{+0.173}_{-1.089}$ & $ 8.610^{+0.157}_{-1.038}$ 
	\\
	
	$\log f_{\rm c}$ & $-1.276^{+0.117}_{-0.205}$ & $-1.188^{+0.142}_{-0.048}$ & $-1.370^{+0.081}_{-0.067}$ & $-1.888^{+0.025}_{-0.144}$  
	\\
	
	$\log (M_{\rm *,c}/\msun)$ & $10.216^{+0.108}_{-0.051}$ & $10.151^{+0.260}_{-0.009}$ & $10.295^{+0.062}_{-0.009}$ & $10.315^{+0.026}_{-0.014}$ 
	\\
	
	$\sigma_{\rm c}$ & $ 0.235^{+0.308}_{-0.027}$ & $ 0.323^{+0.329}_{-0.023}$ & $ 0.287^{+0.064}_{-0.015}$ & $ 0.232^{+0.029}_{-0.013}$
	\\	
	
	$\langle f_{\rm sat}\rangle$ & $ 0.165^{+0.004}_{-0.004}$ & $ 0.155^{+0.003}_{-0.002}$ & $ 0.145^{+0.003}_{-0.003}$ & $ 0.137^{+0.006}_{-0.001}$ 
	\\
	
	$\langle b_{\rm g}\rangle$ & $ 1.103^{+0.057}_{-0.027}$ & $ 1.223^{+0.013}_{-0.026}$ & $ 1.374^{+0.015}_{-0.062}$ & $ 1.420^{+0.025}_{-0.036}$
	
	\enddata	
	\tablecomments{The average satellite fraction
		$\langle f_{\rm sat}\rangle$ and average galaxy bias
		$\langle b_{\rm g}\rangle$ of the ELG samples at each
		redshift interval are also displayed.}
	
\end{deluxetable*}
Following G18, the BOSS LRG is selected from the Data Release 12 of
the BOSS galaxy sample \citep{Reid16}, by applying the additional
color selection of $g-i>2.35$ to remove the blue/star-forming galaxies
as proposed in \cite{Masters11} \citep[see also][]{Maraston13}. The
clustering measurements for BOSS LRGs are presented in two stellar mass
bins of $11<\log(M_*/\msun)<11.5$ and $11.5<\log(M_*/\msun)<12$, while
the SMF is measured in $11<\log(M_*/\msun)<12$ with a bin size of
$\Delta\log M_*=0.1$. We refer the readers to Section 2.1 of G18 for
more details of the BOSS galaxy sample. As a reference, the SMF
measurement of the BOSS LRGs at $0.7<z<0.8$ is also shown in the
lower-left panel of Figure~\ref{fig:wp} as the red circles.
Obviously, the high-mass end of the galaxy SMF is dominated by the
quenched (red) galaxies.

As the BOSS LRGs were observed with completely different target
selections from those of the eBOSS ELGs, and to see whether they have
different $\Phi_{\rm c}^{\rm q}(M_*|M) $ from
$\Phi_{\rm c}^{\rm sf}(M_*|M) $, we use another four parameters
($M_{*,0}^\prime$, $M_1^\prime$, $\alpha^\prime$, $\beta^\prime$) for
the SHMR of the BOSS LRGs as in Equation~(\ref{eq:smhm}) and three
parameters ($f_{\rm c}^\prime$, $M_{\rm *,c}^\prime$,
$\sigma_{\rm c}^\prime$) for the LRG stellar mass completeness
function as in Equation~(\ref{eq:cmstar}). The observed galaxy
clustering and SMF measurements of the BOSS LRGs can be modeled by
replacing $f_{\rm sf}(M)$ with $f_{\rm q}(M)$ in Equations
~(\ref{eq:ncen_int})--(\ref{eq:phisat}).

With all of these data and models available for galaxies in the redshift
interval of $0.7<z<0.8$, we proceed to make our model constraints
using the MCMC method.  The best-fitting quenched fraction $f_{\rm q}(M)$
is shown as the open circles in Figure~\ref{fig:fred}, with the
best-fitting value of $\log(M_{\rm q}/\msun)=12.83\pm0.24$,
which is in good agreement with that obtained in \cite{Tinker13}
(their Figure~9).

Assuming that this quenched fraction $f_{\rm q}(M)$ with the
best-fitting value of $\log(M_{\rm q}/\msun)=12.83$ will not evolve
significantly, which is also supported by the lack of evolution of
$f_{\rm q}(M_*)$ from the literature, in our second step, we proceed to
make our model constraints for the eBOSS ELGs in the redshift range of
$0.8<z<1.2$\footnote{In the future, with the completion of the eBOSS LRG
  program, we will be able to extend the constraints on the quenched
  fraction to these higher redshift bins.}.

We show in Figure~\ref{fig:wp} the best-fitting models using the
corresponding solid lines for $\wprp$ (upper panels) and SMF (lower
panels), respectively. Our best-fitting models show very good
agreement with the measured $\wprp$ at all redshifts and stellar mass
bins. In addition, given the small errors of the SMF measurements, the
good agreement between the two demonstrates that our functional form
of the stellar mass completeness is reasonable.

Finally, the best-fitting model parameters are displayed in
Table~\ref{tab:model}, where the average satellite fraction
$\langle f_{\rm sat}\rangle$ and average galaxy bias
$\langle b_{\rm g}\rangle$ at each redshift interval are also
given. In addition to these fiducial modeling and fitting, we have
also tested some alternatives to quenching and scatter modeling to
check the robustness of our results, which are provided in the
Appendix.

\subsection{Model constraints}\label{sec:constraints}
\begin{figure*}
	\centering
	\includegraphics[width=0.8\textwidth]{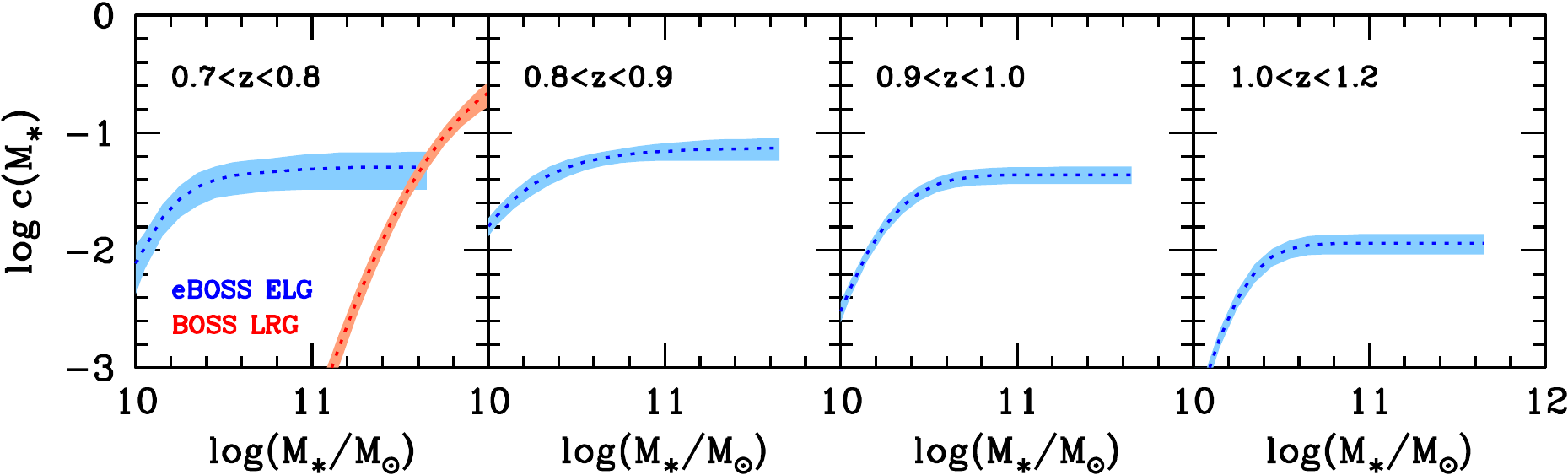}
	\caption{Best-fitting stellar mass completeness functions,
		$c(M_*)$, for the eBOSS ELGs from $z=0.7$ to $z=1.2$ (blue
		dotted line in each panel). The shaded regions are the
		corresponding error distributions. The constraints to the
		completeness function of the BOSS LRGs at $0.7<z<0.8$ is
		also shown as the red dotted line with the shaded area. }
	\label{fig:ncen}
\end{figure*}
\begin{deluxetable*}{lrrrrr}
	\tabletypesize{\scriptsize}	
	\centering	
	\tablecaption{Completeness Functions $\log c(M_*)$ for eBOSS ELGs \label{tab:comp}} 
	\tablehead{$\log (M_*/\msun)$  & $0.7<z<0.8$ & $0.8<z<0.9$ & $0.9<z<1.0$ & $1.0<z<1.2$ }
	\startdata
	10.0 & $-1.977^{+0.129}_{-0.185}$  & $-1.707^{+0.065}_{-0.077}$  & $-2.331^{+0.064}_{-0.076}$  & $-3.204^{+0.082}_{-0.101}$ \\
	
	10.2 & $-1.563^{+0.114}_{-0.154}$  & $-1.445^{+0.072}_{-0.087}$  & $-1.789^{+0.059}_{-0.069}$  & $-2.415^{+0.068}_{-0.080}$ \\
	
	10.4 & $-1.399^{+0.114}_{-0.155}$  & $-1.293^{+0.065}_{-0.077}$  & $-1.506^{+0.060}_{-0.069}$  & $-2.054^{+0.069}_{-0.083}$ \\
	
	10.6 & $-1.349^{+0.118}_{-0.163}$  & $-1.215^{+0.056}_{-0.064}$  & $-1.396^{+0.059}_{-0.069}$  & $-1.954^{+0.073}_{-0.088}$ \\
	
	10.8 & $-1.323^{+0.121}_{-0.168}$  & $-1.175^{+0.060}_{-0.070}$  & $-1.364^{+0.065}_{-0.076}$  & $-1.940^{+0.076}_{-0.092}$ \\
	
	11.0 & $-1.303^{+0.126}_{-0.177}$  & $-1.152^{+0.071}_{-0.085}$  & $-1.358^{+0.067}_{-0.080}$  & $-1.940^{+0.076}_{-0.092}$ \\
	
	11.2 & $-1.294^{+0.130}_{-0.187}$  & $-1.140^{+0.078}_{-0.096}$  & $-1.357^{+0.068}_{-0.081}$  & $-1.940^{+0.076}_{-0.092}$ \\
	
	11.4 & $-1.294^{+0.130}_{-0.187}$  & $-1.134^{+0.082}_{-0.101}$  & $-1.357^{+0.068}_{-0.081}$  & $-1.940^{+0.076}_{-0.092}$ \\
	
	11.6 & $-1.291^{+0.131}_{-0.189}$  & $-1.132^{+0.084}_{-0.104}$  & $-1.357^{+0.068}_{-0.081}$  & $-1.940^{+0.076}_{-0.092}$
	\enddata 
	\tablecomments{The measurements are shown for the completeness
		function, $\log c(M_*)$, of the whole ELG samples at
		different redshifts, as defined in Eq.~(\ref{eq:cmstar}).}
\end{deluxetable*}
After we constrain our models as outlined in Section
\ref{sec:fitting}, we provide the related model constraints in this
subsection. 

\subsubsection{Stellar Mass Completeness}
We show in Figure~\ref{fig:ncen} the best-fitting stellar mass
completeness functions at different redshifts as the blue dotted lines,
with the shaded regions as the 1$\sigma$ error distributions. As seen
from the figure, as a result of the complicated target selections, the
eBOSS ELG sample is very incomplete, with the average high-mass end
completeness varying from about 1\% to 10\% depending on the
redshift. The BOSS LRG sample completeness at $0.7<z<0.8$ is shown as
the red dotted line with the shaded area in the leftmost panel. The BOSS
LRG sample is more complete at the massive end where they dominate the
galaxy SMF.

The ELG samples at lower redshifts are relatively more complete than
the higher redshift ones, as the target selection cuts are designed to
choose galaxies in the redshift range of $0.7<z<1.1$ \citep[see Figure~4 of][]{Raichoor17}. The completeness
of the ELG samples decreases significantly toward the low-mass
end. For galaxies with stellar mass around $10^{10}\msun$, the eBOSS
ELG sample only consists of less than 1\% of the star-forming galaxy
population.  The values of the stellar mass completeness for eBOSS
ELGs are listed in Table~\ref{tab:comp}. We caution that the low
completeness of the eBOSS ELGs makes them less representative of the
entire star-forming galaxy population, which can be improved with the
next-generation ELG surveys with higher completenesses as in DESI \citep[see
e.g.,][]{Gonzalez-Perez18}.

Thanks to the large sky coverage of the eBOSS ELG program, as shown in Table~\ref{tab:data}, we have a reasonable number of galaxies at $z\sim1$ to achieve accurate clustering measurements, despite of the low sampling rates. While the clustering measurements only weakly depend on the sample completeness as shown in Equations (\ref{eq:xicc_sim}) and (\ref{eq:nc}), the remaining question is whether the observed \OII\ emitters are a representative subsample of the overall star-forming galaxy population. By comparing to the SFR vs. stellar mass relation in Figure~1 of \cite{Moustakas13} at similar redshifts, we find that the eBOSS ELG sample is selecting typical star-forming galaxies that lie within the star-formation sequence, with the mean $\log(\rm{SFR}/\msun\,{\rm yr}^{-1})$ at $0.65<z<0.8$ and $0.8<z<1.0$ being $0.89\pm0.42$ and $1.22\pm0.45$, respectively. As will be shown in \S\ref{sec:intrinsicSMF}, it is also encouraging that the recovered intrinsic galaxy SMFs are in good agreement with the literature, further confirming that the eBOSS ELGs are in general representative of the star-forming populations at the corresponding redshifts.

\subsubsection{Stellar-Halo Mass Relation}
\begin{figure*}
	\centering
	\includegraphics[width=0.8\textwidth]{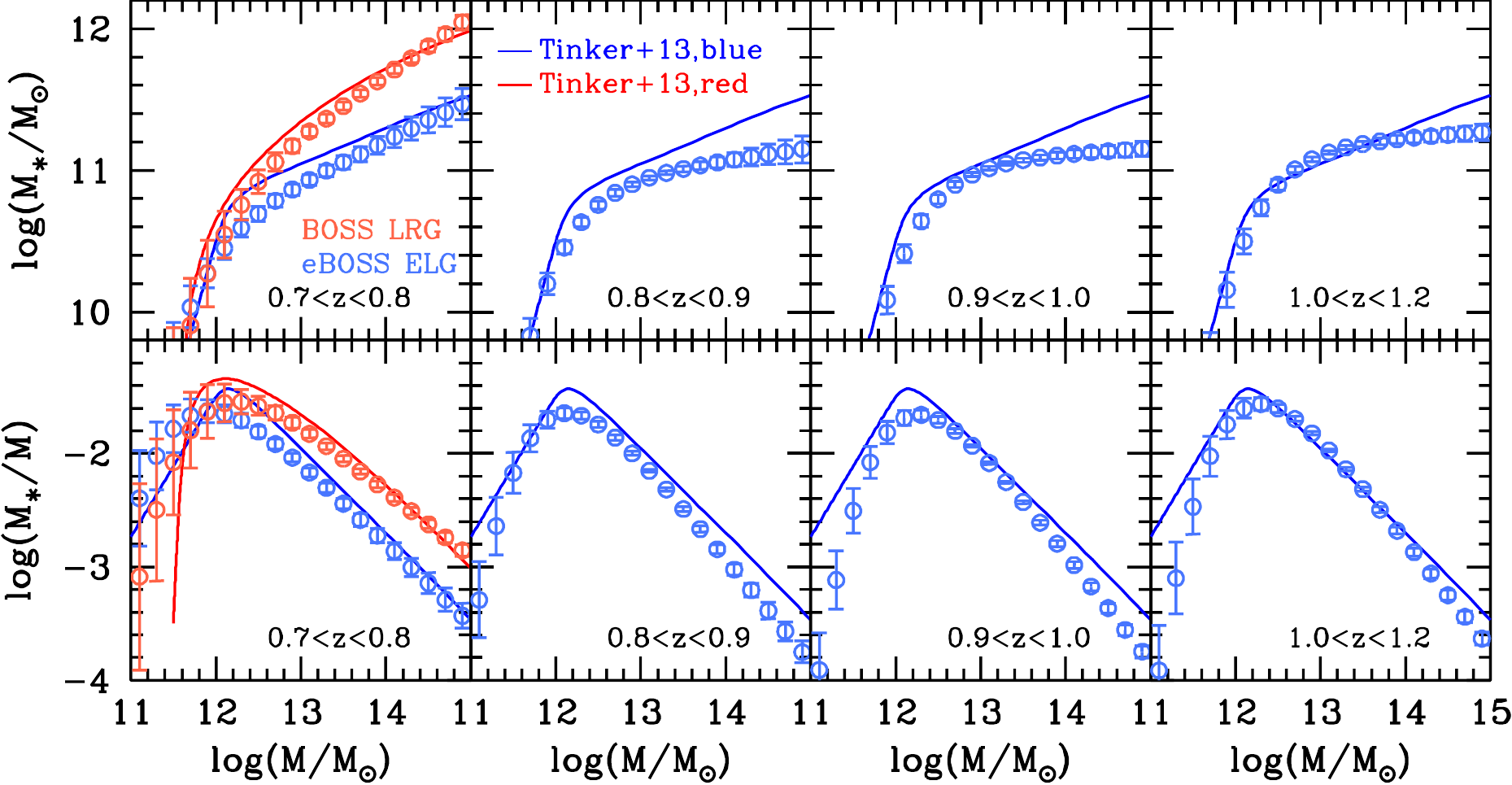}
	\caption{Best-fitting stellar-halo mass relations (top panels)
		and the stellar-to-halo mass ratios (bottom panels). The
		blue open circles are for the eBOSS ELGs, with the red open
		circles for the BOSS LRG at $0.7<z<0.8$. The model results
		of \cite{Tinker13} for red and blue galaxies (see text) are
		also shown for comparison as the red and blue solid lines.}
	\label{fig:smhm}
\end{figure*}

We show in the top panels of Figure~\ref{fig:smhm} the predicted SHMRs
of ELGs as the blue circles with errors. The SHMR of the BOSS LRG in
$0.7<z<0.8$ is shown as the red circles in the left most panel. There
is only weak evolution in the shape of the ELG SHMR from $z=1.2$ to
$z=0.7$ for halos of $\log (M/\msun)<13$. The high-mass end slope $\alpha$ of the SHMR is around 0.35 at $0.7<z<0.8$, while it becomes very flat ($\alpha\sim0$) for higher redshift samples. It implies that there are only very few star-forming galaxies of $M_*>10^{11.2}\msun$ at these redshifts. But the SMFs of the star-forming galaxies at the massive end are still not vanishing due to the large scatter $\sigma_*$ for the high redshift samples. However, we caution that the error on the best-fitting slope $\alpha$ may be underestimated in our model, as it is mainly constrained by the observed SMFs. The constraining power of the clustering measurements is weakened by the low S/N.   

By comparing the SHMRs between the eBOSS ELG and BOSS LRG samples at
$0.7<z<0.8$, we find that the ELGs and LRGs have very similar SHMRs
for halos of $\log(M/\msun)<12$. For more massive halos, there are
significant differences between the two SHMRs, with the quiescent
galaxies having a much steeper slope at the massive end as also shown
in Figure~9 of G18. \cite{Tinker13} have constrained the SHMRs for the
quiescent and star-forming galaxies over the redshift range of
$0.2<z<1$ by using measurements of the galaxy angular clustering and
galaxy--galaxy lensing in the COSMOS field. For fair comparisons, we
also show the results of \cite{Tinker13} at $0.74<z<1$ as the red and
blue solid lines for the quiescent and star-forming populations,
respectively. The trend in the discrepancy between the two SHMRs is in good
agreement with those of \cite{Tinker13}. They attributed the
differences to the much larger scatter $\sigma_*$ of the star-forming
galaxies in their models. However, in our model, we assume the same
scatter $\sigma_*$ for both the star-forming and quiescent
galaxies. Therefore, the differences in the SHMRs may reflect the
intrinsic variation in the average galaxy stellar mass at a given halo
mass for the two populations.

We note that the high mass end of the galaxy SMF is dominated by the
quiescent galaxies. Since most previous results of the galaxy SHMR
constraints come from fitting the overall galaxy SMFs \citep[see
e.g.,][]{Yang12,Behroozi13,Moster13,Rodriguez-Puebla17}, they are not
directly comparable to our model predictions of the ELG SHMR at these
redshifts.
\begin{figure*}
	\centering
	\includegraphics[width=0.6\textwidth]{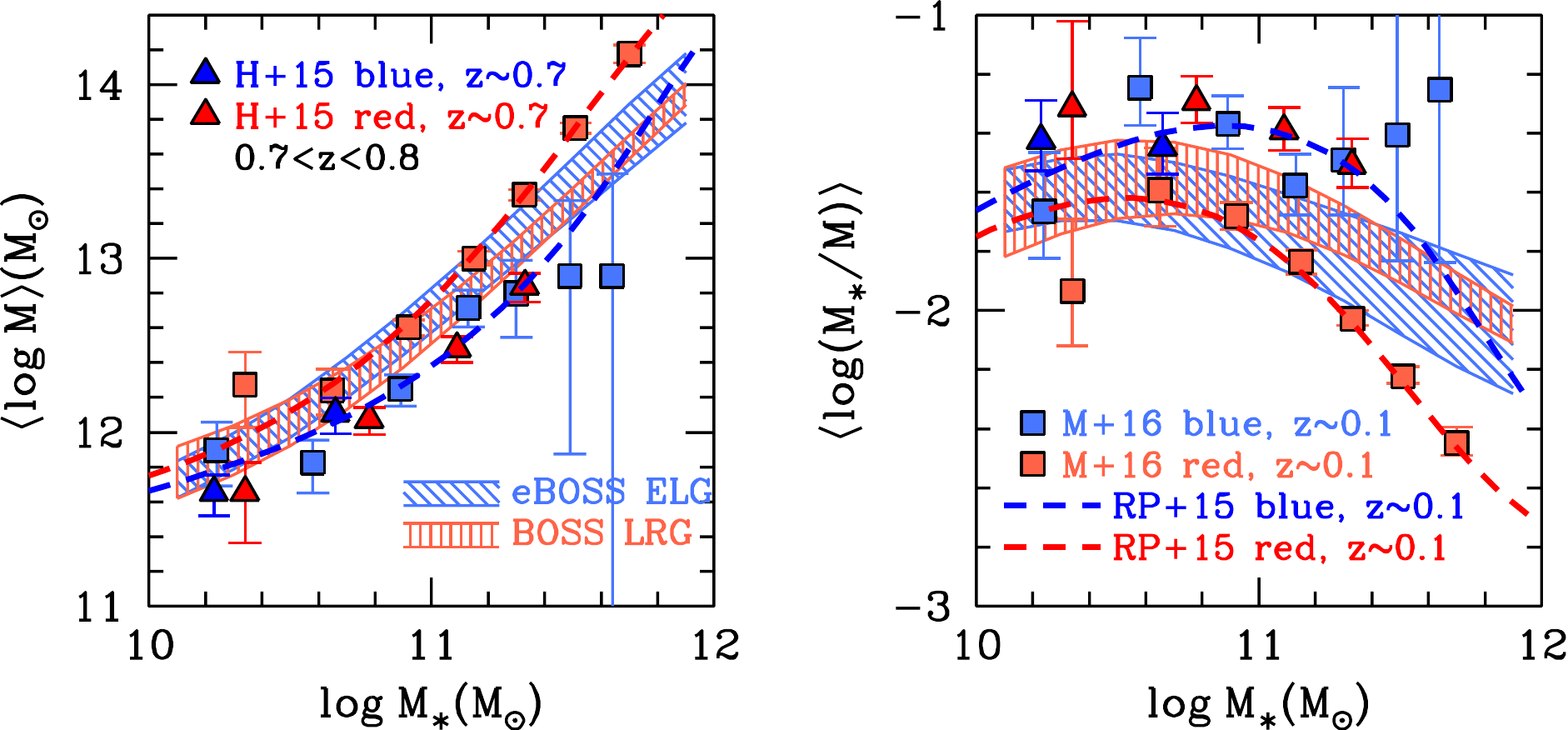}
	\caption{Left: the model constraints to the average halo mass at a given stellar mass. The 1$\sigma$ ranges of our best-fitting models for the BOSS LRGs and eBOSS ELGs at $0.7<z<0.8$ are shown as the red and blue shaded regions, respectively. The weak lensing measurements from \cite{Hudson15} (denoted as ``H+15'') at $z\sim0.7$ and \cite{Mandelbaum16}  (denoted as ``M+16'') at $z\sim0$ are shown as the filled triangles and squares, respectively. We also display the halo model predictions from \cite{Rodriguez-Puebla15}  (denoted as ``RP+15'') at $z\sim0$ for comparison. Right: the corresponding stellar-to-halo mass ratios of different measurements in the left panel.}
	\label{fig:hmsm}
\end{figure*}

We show in the bottom panels of Figure~\ref{fig:smhm} the
stellar-to-halo mass ratios as a function of the halo mass at
different redshifts. The peaks of the stellar-to-halo mass ratio
happen at around $M\sim10^{12}\msun$, consistent with the findings in
the literature \citep[see e.g.,][]{Yang12,Behroozi13}. It is
interesting that the peak locations of the stellar-to-halo mass ratios
for the quiescent and star-forming galaxies are roughly coincident
with each other. It implies that the transition from the star-forming
to quiescent galaxies might be most efficient in halos of
$M\sim10^{12}\msun$.

To appropriately compare with the observational constraints from the weak lensing measurements, we also calculate the average halo masses at given stellar masses for star-forming and quiescent central galaxies as \citep{Hudson15,Rodriguez-Puebla15},
\begin{eqnarray}
\langle M|M_*\rangle_{\rm sf}&=&\frac{\int \Phi_{\rm c}^{\rm sf}(M_*|M)f_{\rm sf}(M)n_{\rm h}(M) M dM}{\int \Phi_{\rm c}^{\rm sf}(M_*|M)f_{\rm sf}(M)n_{\rm h}(M) dM}\\
\langle M|M_*\rangle_{\rm q}&=&\frac{\int \Phi_{\rm c}^{\rm q}(M_*|M)f_{\rm q}(M)n_{\rm h}(M) M dM}{\int \Phi_{\rm c}^{\rm q}(M_*|M)f_{\rm q}(M)n_{\rm h}(M) dM},
\end{eqnarray}
which are significantly different from the SHMR of $\langle M_*|M\rangle$ due to the existence of scatter $\sigma_*$, especially at the massive end \citep{Behroozi10,Tinker13}. As we only have the measurements of both LRGs and ELGs at $0.7<z<0.8$, we show our best-fitting model constraints to $\langle M|M_*\rangle_{\rm sf}$ and $\langle M|M_*\rangle_{\rm q}$ in the left panel of in Figure~\ref{fig:hmsm}, while the stellar-to-halo mass ratios for the two samples as a function of the stellar mass are shown in the right panel. For comparison, we also display the galaxy-galaxy weak lensing measurements for red and blue galaxies from \cite{Hudson15} at $z\sim0.7$ and from \cite{Mandelbaum16} at $z\sim0$. The halo model constraints from \cite{Rodriguez-Puebla15} at $z\sim0$ are also shown as the dotted lines. 

Our model predictions of the $\langle M|M_*\rangle$ for star-forming and quiescent galaxies almost overlap with each other over the stellar mass range of $10<\log(M_*/\msun)<12$ at $0.7<z<0.8$, consistent with the results of \cite{Tinker13} (their Figure~7). Compared to the weak lensing measurements from \cite{Hudson15}, our model predictions tend to be slightly higher. However, as noted by \cite{Coupon15} (their Figure~11), the stellar mass measurements of \cite{Hudson15} are likely biased high due to the lack of NIR data and the estimated halo mass is dependent on the details of modeling the contributions of subhalos in the lensing signals. The low redshift measurements of \cite{Mandelbaum16} and \cite{Rodriguez-Puebla15} agree well with each other. They found apparent differences in the halo masses for red and blue galaxies with the same stellar masses, considering the small errors in the red galaxy measurements. Although such a trend is not shown in our best-fitting model constraints at $z\sim0.7$, the overall evolution of $\langle M|M_*\rangle$ is generally weak, especially for galaxies with $\log(M_*/\msun)<11$. As pointed out by \cite{Lapi18a}, the weak evolution of the mass ratio $\langle M_*/M\rangle$ for low mass galaxies indicates the joint effects of the star formation and dark matter accretion along the cosmic time, which would have important constraints on the galaxy formation and evolution models, while the evolution of more massive galaxies may be quite different \citep{Shankar14,Bernardi16,Lapi18b}.

\subsection{Model predictions}\label{sec:predictions}
\begin{figure*}
	\centering
	\includegraphics[width=1.0\textwidth]{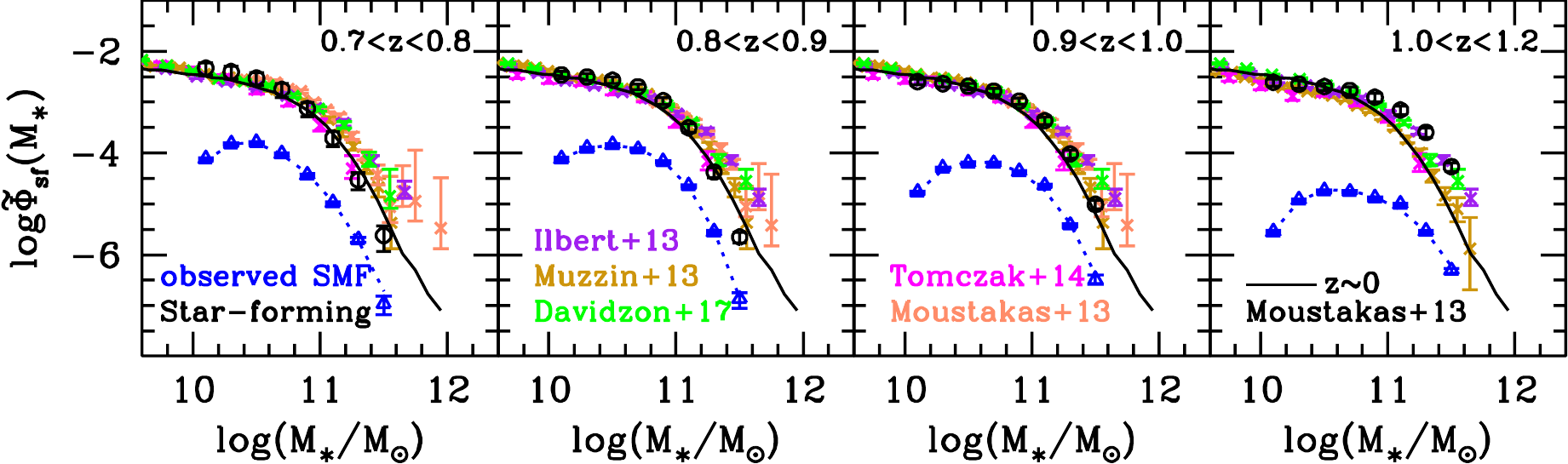}
	\caption{Intrinsic stellar mass functions for the star-forming
		galaxies. The black circles are our best-fitting models,
		while crosses of different colors represent the previous
		measurements of \cite{Ilbert13}, \cite{Muzzin13},
		\cite{Moustakas13}, \cite{Tomczak14} and \cite{Davidzon17}. 
		We note that the measurements of previous literature are in much larger redshift bins compared to our results. 
			They are repeated in some neighboring bins for simple comparisons. The measurements
		at the low redshifts of $z\sim0$ from \cite{Moustakas13} are
		also shown as the black solid line in each panel for
		comparisons. The dotted line in each panel shows the
		observed eBOSS ELG SMF.}
	\label{fig:smftblue}
\end{figure*}
With the above model constraints, as we outlined in Section
\ref{subsec:predictions}, it is quite straightforward to make some
model predictions, such as the intrinsic galaxy SMFs, the HOD, the
galaxy bias and the satellite fraction.

\subsubsection{Intrinsic Galaxy SMFs}\label{sec:intrinsicSMF}
We show in Figure~\ref{fig:smftblue} the predicted intrinsic galaxy
SMFs as black open circles for the star-forming galaxies from our
best-fitting models. The measurements of the intrinsic SMFs for the
star-forming galaxies are also displayed in Table~\ref{tab:smft}. We
show for comparison the previous measurements in the literature at the
corresponding redshifts, including \cite{Ilbert13},
\cite{Moustakas13}, \cite{Muzzin13}, \cite{Tomczak14} and \cite{Davidzon17} as crosses of
different colors. The local-universe measurements of
\cite{Moustakas13} at $z\sim0$ are shown as the solid line in each
panel to see the evolution effect. We have corrected the different
assumptions of the IMF, SPS models and dust attenuation laws in the
literature following \cite{Rodriguez-Puebla17}, to be consistent with
our model assumptions. The blue open triangles are the observed galaxy
SMFs as in Figure~\ref{fig:wp}, with the dotted lines showing the
best-fitting models.

Our best-fitting models for the intrinsic galaxy SMFs of the star-forming
galaxies are generally in good agreement with the literature. While at
the low-mass end ($M_*<10^{10}\msun$) the SMFs from the previous
literature are roughly consistent with each other, there are larger
discrepancies for more massive galaxies. As those previous
measurements are made with photometric redshifts covering small but
deep survey area, the variations could be caused by the sample
variance effect due to the limited volumes and also other systematic
effects, such as the different methods to estimate the galaxy stellar
mass and to discriminate between the star-forming (blue) and quiescent
(red) galaxies.
\begin{deluxetable*}{lllll}
	\tabletypesize{\scriptsize}
	\centering	
	\tablecaption{Intrinsic SMF $\log\tilde{\Phi}_{\rm sf}(M_*)$ For Star-Forming Galaxies \label{tab:smft}} 
	\tablehead{$\log (M_*/\msun)$ & $0.7<z<0.8$ &$0.8<z<0.9$& $0.9<z<1.0$ & $1.0<z<1.2$}
	\startdata
10.1 & $-2.330^{+0.108}_{-0.145}$  & $-2.462^{+0.065}_{-0.077}$  & $-2.598^{+0.054}_{-0.062}$  & $-2.610^{+0.066}_{-0.077}$ \\

10.3 & $-2.404^{+0.108}_{-0.144}$  & $-2.503^{+0.063}_{-0.074}$  & $-2.635^{+0.055}_{-0.063}$  & $-2.646^{+0.066}_{-0.078}$ \\

10.5 & $-2.527^{+0.110}_{-0.149}$  & $-2.568^{+0.059}_{-0.069}$  & $-2.689^{+0.056}_{-0.065}$  & $-2.694^{+0.068}_{-0.081}$ \\

10.7 & $-2.749^{+0.114}_{-0.156}$  & $-2.698^{+0.055}_{-0.063}$  & $-2.785^{+0.058}_{-0.067}$  & $-2.768^{+0.071}_{-0.085}$ \\

10.9 & $-3.124^{+0.117}_{-0.161}$  & $-2.975^{+0.060}_{-0.069}$  & $-2.980^{+0.063}_{-0.074}$  & $-2.903^{+0.075}_{-0.091}$ \\

11.1 & $-3.705^{+0.123}_{-0.171}$  & $-3.503^{+0.073}_{-0.088}$  & $-3.362^{+0.067}_{-0.080}$  & $-3.156^{+0.077}_{-0.094}$ \\

11.3 & $-4.529^{+0.136}_{-0.199}$  & $-4.374^{+0.084}_{-0.104}$  & $-4.017^{+0.072}_{-0.086}$  & $-3.593^{+0.078}_{-0.096}$ \\

11.5 & $-5.619^{+0.183}_{-0.321}$  & $-5.650^{+0.092}_{-0.116}$  & $-5.010^{+0.076}_{-0.093}$  & $-4.267^{+0.078}_{-0.095}$
	\enddata   						
	
	\tablecomments{The stellar mass function measurements are in units of $\rm{Mpc}^{-3}\rm{dex}^{-1}$. }
\end{deluxetable*}

\begin{deluxetable*}{cccc}
	\tabletypesize{\scriptsize}	
	\centering	
	\tablecaption{Best-fit Schechter Function Parameters for the ELGs \label{tab:schechter}} 
	\tablehead{redshift range & $\log(\Phi^\ast/\rm{Mpc}^{-3}\rm{dex}^{-1})$ & $\log(M^\ast_{\rm c}/\msun)$ & $\alpha^\ast$}
	\startdata
	
	$0.7<z<0.8$ & $-2.658\pm0.162$ & $10.637\pm0.077$ & $-1.253\pm0.328$ \\
	
	$0.8<z<0.9$ & $-2.559\pm0.039$ & $10.585\pm0.020$ & $-0.791\pm0.123$ \\
	
	$0.9<z<1.0$ & $-2.793\pm0.043$ & $10.735\pm0.018$ & $-0.936\pm0.095$ \\
	
	$1.0<z<1.2$ & $-2.939\pm0.059$ & $10.938\pm0.024$ & $-1.071\pm0.096$  
	
	\enddata 
\end{deluxetable*}

\begin{figure*}
	\centering
	\includegraphics[width=0.8\textwidth]{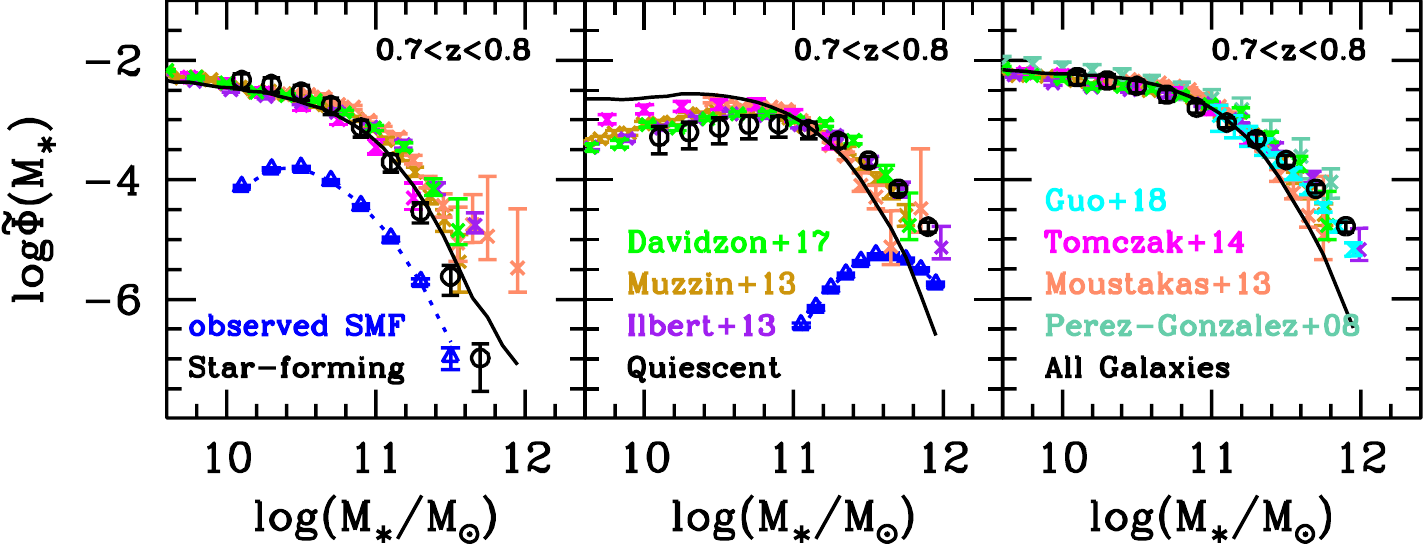}
	\caption{Intrinsic stellar mass functions for the star-forming
		galaxies (left panel), quiescent galaxies (middle panel) and
		all galaxies (right panel) at $0.7<z<0.8$. The black circles
		are our best-fitting models, while crosses of different
		colors represent the measurements of
		\cite{Perez-Gonzalez08}, \cite{Ilbert13}, \cite{Muzzin13},
		\cite{Moustakas13}, \cite{Tomczak14}, \cite{Davidzon17} and G18. The black
		solid lines are the measurements at $z\sim0$ from
		\cite{Moustakas13}. The open triangles with the dotted lines
		are the observed eBOSS ELG SMF and the corresponding
		best-fitting models. For comparison purposes, we have extended
		our predicted galaxy SMFs over the whole stellar mass range
		of $10^{10}\msun<M_*<10^{12}\msun$.}
	\label{fig:smftred}
\end{figure*}
As the eBOSS ELG sample covers a significantly larger volume than the
previous surveys, it suffers less from the sample variance
effect. In addition, the star-forming galaxies are identified with the
\OII\ emission line, which is more reliable than the discrimination
method with the color-magnitude diagram used in previous
surveys. However, we note that the stellar masses of the eBOSS ELGs
are obtained with a small number of available bands in DECaLs and
WISE. The stellar mass estimates from the SED fittings are therefore
less accurate compared to the previous deep surveys with broadband
photometry.

For the star-forming galaxy population, there is only weak evolution
of the SMF at $M_*<10^{11}\msun$ from $z=1.2$ to $0.7$. Our
best-fitting models predict that the number density of ELGs with
$M_*>10^{11}\msun$ decreases significantly around $z\sim1$. However, we caution that our model predictions at $z>1$ might be affected by the low sample selection rates ($\sim1\%$) which make the observed ELGs less representative of the overall star-forming galaxy population. Compared to the SMF measurements of the star-forming galaxies
at $z\sim0$, it seems that there is almost no evolution from $z\sim0$
to $z=0.9$, which is consistent with the conclusions of previous
measurements as in \cite{Ilbert13}, \cite{Muzzin13} and \cite{Lapi17}.

To quantitatively compare with the corresponding measurements in the
literature, we also fit the intrinsic SMFs for the star-forming
galaxies with the standard single Schechter function
\citep{Schechter76},
\begin{equation}
  \Phi(M_*)=(\ln 10)\Phi^{\ast} \exp\left(-\frac{M_*}{M^{\ast}_{\rm c}}\right)\left(\frac{M_*}{M^{\ast}_{\rm c}}\right)^{1+\alpha^\ast}
\end{equation}
The best-fit parameters $\Phi^{\ast}$, $M^{\ast}_{\rm c}$ and
$\alpha^\ast$ are shown in Table~\ref{tab:schechter}. 

As we have the measurements for both the quiescent and star-forming
galaxies at $0.7<z<0.8$, we can predict the total galaxy SMF at this
redshift and compare it with that in the literature to check the performance of
the best-fitting model. We show in Figure~\ref{fig:smftred} the
intrinsic SMFs for the star-forming galaxies (left panel), quiescent
galaxies (middle panel), and all galaxies (right panel) at
$0.7<z<0.8$. As in Figure~\ref{fig:smftblue}, we also show for
comparison the various measurements from the literature, including
those from \cite{Perez-Gonzalez08}, \cite{Ilbert13}, \cite{Muzzin13},
\cite{Moustakas13}, \cite{Tomczak14} and G18. For comparison purposes,
we have extended our predicted galaxy SMFs over the whole stellar mass
range of $10^{10}\msun<M_*<10^{12}\msun$, although the observed
quiescent and star-forming galaxy SMFs (shown as the blue open
triangles) are only limited to smaller stellar mass ranges. The
corresponding best-fitting models to the observed SMFs are displayed
as the dotted lines.

Our measurements of the total galaxy SMF show good agreement with that
of G18, as we are using the same set of BOSS LRG data. In general, our
measurements of the intrinsic SMFs are in agreement with those from
the literature. There are slightly larger discrepancies in
the different measurements of the quiescent galaxy SMF at
$M_*<10^{11}\msun$. Although our measurements are simply extensions of
the models for the more massive BOSS LRGs, it is hard to justify the
discrepancies. But the total galaxy SMFs from our model and those from the
literature tend to be consistent with each other. 

Comparing to the
galaxy SMF measurements at $z\sim0$ from \cite{Moustakas13} (solid
black lines), it seems that there is significant evolution in the SMF
of the quiescent galaxies from $z\sim0$ to $z=0.8$. A detailed study of
the evolution of the galaxy SMF, as well as the star formation
processes, will be presented in the future work. This figure implies
that our method is a powerful way of reconstructing the galaxy
intrinsic SMFs with incomplete survey samples and studying the process of
galaxy quenching. Measurements of both the quiescent and star-forming
galaxies at the same redshifts (as in e.g., DESI) would help further
constrain the galaxy SMFs of different populations over much larger
redshift ranges.

\subsubsection{Halo Occupation Distribution}

\begin{figure*}
	\centering
	\includegraphics[width=0.8\textwidth]{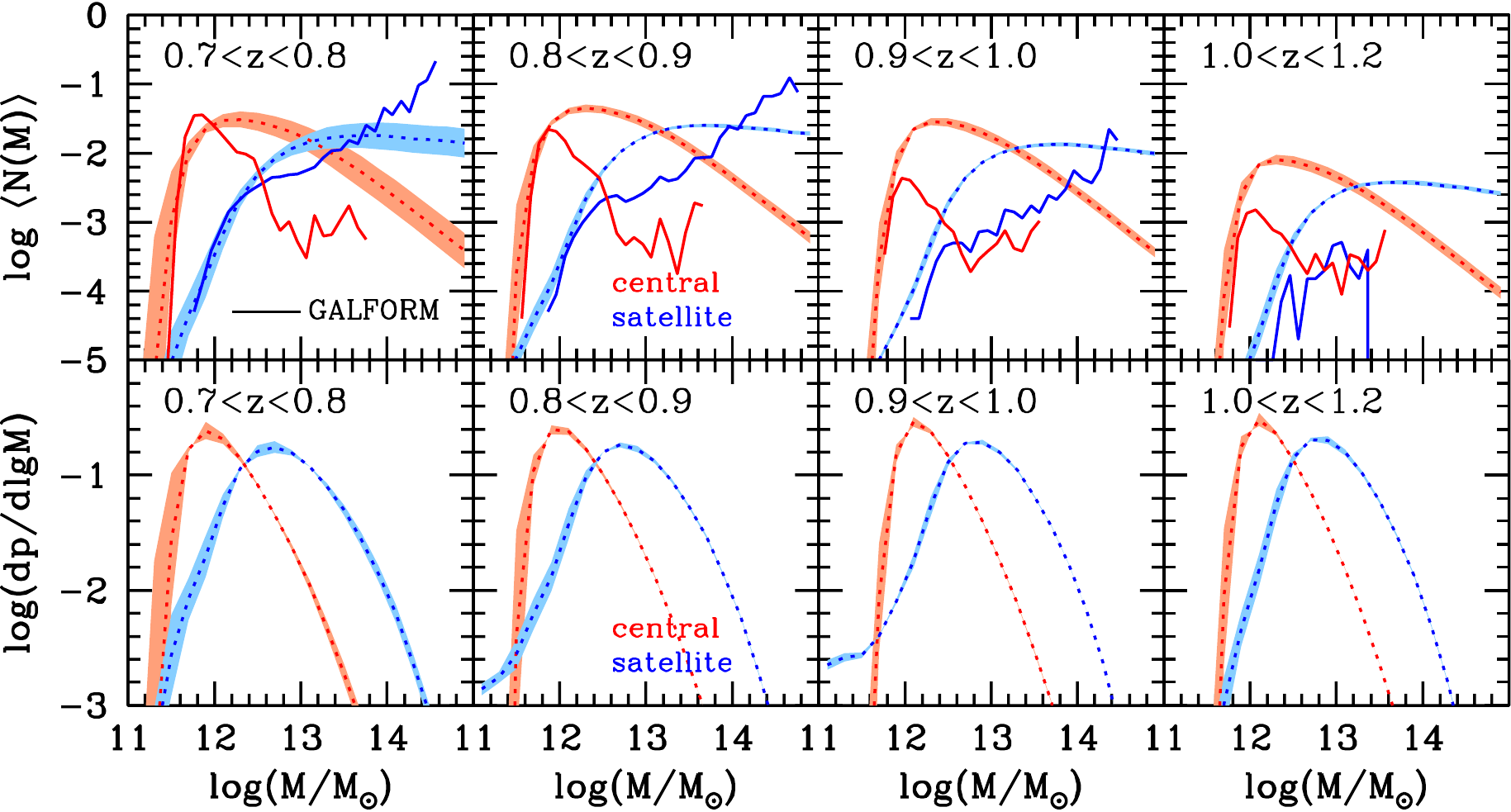}
	\caption{Halo occupation distribution functions for the
          central (red dotted lines) and satellite galaxies (blue
          dotted lines). The shaded area are the 1$\sigma$
          distribution. For comparison, the predictions from the SAM
          of GALFORM are shown as the red and blue solid lines for
          central and satellite galaxies, respectively. }
	\label{fig:hod}
\end{figure*}

We show the best-fitting halo occupation distribution function
$\langle N(M)\rangle$ for the observed ELGs in different redshift bins
in the top panels of Figure~\ref{fig:hod}.  The central and satellite
galaxies are shown with red dotted lines and blue dotted lines,
respectively.  The shaded area represents the 1$\sigma$ error
distribution. The occupation functions differ from the standard HOD
form of \cite{Zheng07} with the significant decrease of occupation
numbers at the massive end \citep[see also][]{Geach12,
  Contreras13}. Because we are only including the star-forming
galaxies that have additional dependence on the quenched fraction
$f_{\rm q}(M)$, the massive halos are dominated by the red/quiescent
central galaxies. In our best-fitting models, the satellite occupation
functions become flat at the massive end, due to the lack of
star-forming satellite galaxies in very massive halos.

\begin{figure*}
	\centering
	\includegraphics[width=0.8\textwidth]{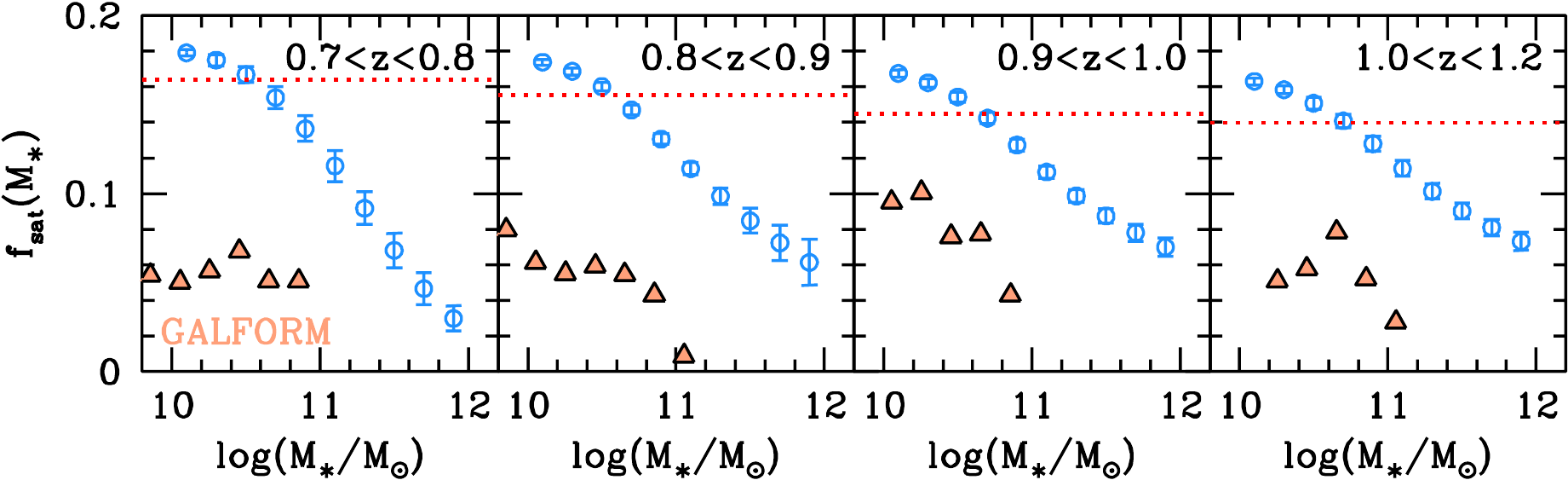}
	\caption{Satellite galaxy fraction $f_{\rm sat}(M_*)$ (open
          circles) from the best-fitting models. The average satellite
          fraction $\langle f_{\rm sat}\rangle$ of each sample is
          displayed as the red dotted line. The model predictions of the SAM of GALFORM are shown as the filled triangles.}
	\label{fig:fsat}
\end{figure*}

\begin{figure*}
\centering
	\includegraphics[width=0.8\textwidth]{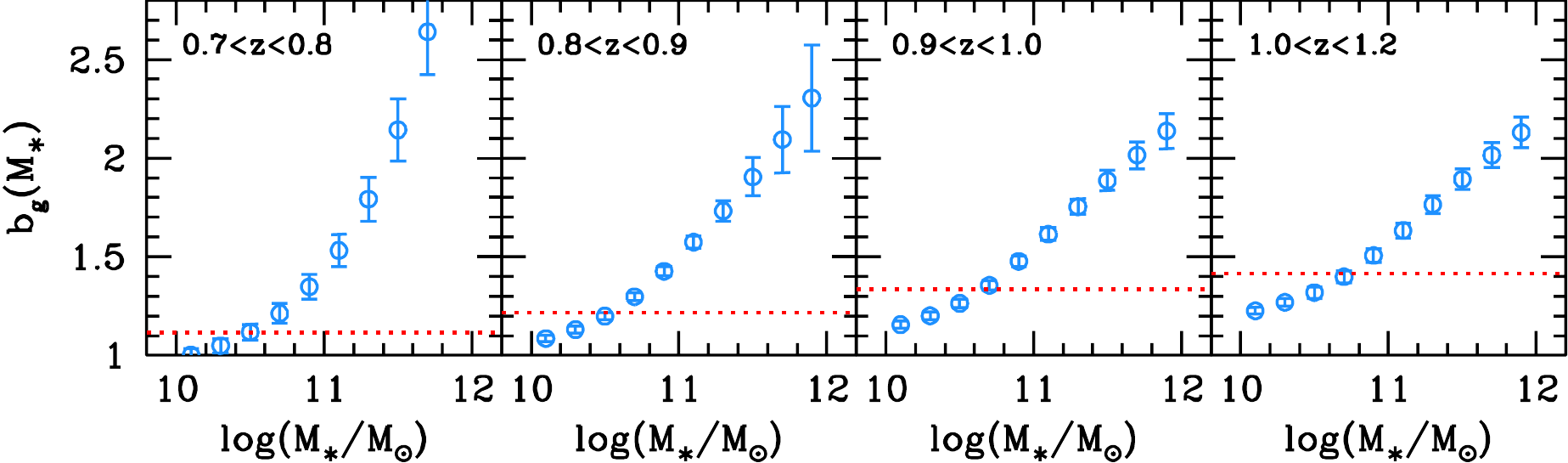}
	\caption{Similar to Figure~\ref{fig:fsat}, but for the galaxy
          bias parameter.}
\label{fig:bias}
\end{figure*}

We also compare our best-fitting models to the predictions from the
semi-analytical model (SAM) of GALFORM with updated model treatments of
\cite{Gonzalez-Perez18} (hereafter GP18) as described in
\cite{Griffin18} (also V. Gonzalez-Perez et al., in prep.), which are
shown as the red and blue solid lines for central and satellite
galaxies, respectively. The galaxies from the SAM of GP18 were
selected using flux, magnitude and color cuts to mimic the eBOSS ELG
target sample from \cite{Raichoor17}. Our predictions show good
agreement with GP18 for the low-mass cutoff profiles, i.e., the ELGs
usually live in halos more massive than $10^{11.4}\msun$. It sets the
requirement of the simulation resolution to generate the mock galaxy
samples for the observed ELGs \citep[see
e.g.,][]{Chuang15,Lippich18}. However, the central galaxy occupation
functions of GP18 are significantly smaller than our model predictions
for massive halos, which may be due to the treatment of the dust attenuation for the most massive star-forming galaxies in the SAM \citep{Gonzalez-Perez18}. 

We also note that the predicted number densities of
the \OII\ emitters in the GALFORM model are significantly smaller than
the observed ones in the eBOSS sample at $z>0.8$ \citep[see Figure~5
of][]{Gonzalez-Perez18}. Moreover, the satellite
fraction for the ELGs in the GP18 model is only around 5\%, implying
that the star formation in the satellite galaxies is suppressed too
effectively. Interestingly, the discrepancies of the satellite occupation numbers in massive halos could possibly be related to the problem of the merger-driven formation model for massive galaxies. As shown in recent studies \citep[e.g.][]{Lapi18b}, the in-situ processes may be more relevant in driving the stellar and black hole mass growth. While our modeling results provide useful constraints to
the  ELGs in the SAMs, detailed comparisons to the SAMs is beyond
the scope of this work.

We show in the bottom panels of Figure~\ref{fig:hod} the normalized
probability distributions of the host halo masses of the central (red
lines) and satellite (blue lines) ELGs, which are generated from the
product of the occupation function and the differential halo mass
function \citep{Zheng09,Guo14}. The peak halo mass distribution of the
central ELGs is around $10^{12}\msun$, while the satellite galaxies
live in more massive halos of $10^{12.7}\msun$. There is only a weak trend
in the evolution of the halo mass distribution at $0.7<z<1.2$,
consistent with the trend in the predicted SHMRs of
Figure~\ref{fig:smhm}.

Although our best-fitting models of the occupation functions have
small scatters, it is related to the fact that we have
assumed a specific functional form of $f_{\rm q}(M)$. We
note that the exact shape of the occupation function at the massive
end is slightly dependent on the functional form of $f_{\rm q}(M)$, as
will be discussed in the Appendix. However, the detailed
high-mass end shape of the occupation function does not affect our
constraints on the host halo mass distributions of the ELGs, as the
halo mass function decreases very fast toward the massive end.

\subsubsection{Satellite Galaxy Fraction}

We show the satellite galaxy fraction in the observed ELG samples as
the open circles in Figure~\ref{fig:fsat}, where the model predictions from the SAM of GALFORM are shown as the filled triangles. The average satellite
fraction $\langle f_{\rm sat}\rangle$ of each sample is displayed as
the red dotted line. The average satellite fraction varies from about
13\% to 17\%, with the higher redshift samples having slightly smaller
$f_{\rm sat}$. The satellite fraction $f_{\rm sat}(M_*)$ is generally
decreasing with the stellar mass. The SAM of GALFORM generally underestimates the satellite fraction of the ELGs at different stellar masses. Since the majority of the observed
ELGs are central galaxies, the discrepancies of the satellite
occupation functions between our models and those of GALFORM in
Figure~\ref{fig:hod} would not have a significant effect on the
predicted sample number densities.

\cite{Favole16} found a satellite fraction of around $22\%$ for a
g-band selected galaxy sample in $0.6<z<1$ using a modified subhalo
abundance matching model to include the incompleteness effect. Their
result is slightly larger than our estimates, but the detailed value
of the satellite fraction is sample-dependent, as their photometrically selected sample may include more satellite galaxies. 

\subsubsection{Galaxy Bias}

The galaxy bias $b_{\rm g}(M_*)$ is shown as the filled circles in
Figure~\ref{fig:bias}, where the average galaxy bias
$\langle b_{\rm g}\rangle$ is also displayed as the dotted lines. The
galaxy bias is apparently a strongly increasing function of the galaxy
stellar mass, varying from $1$ to $3$. The average galaxy bias
$\langle b_{\rm g}\rangle$ increases from $1.1$ at $z\sim0.7$ to $1.4$
at $z\sim1.2$. It is also consistent with the fact that the majority of
the central ELGs in these redshift ranges live in halos of
$\sim10^{12}\msun$ that have similar halo bias values.

\section{Discussions} \label{sec:discussion}

The ICSMF model is a powerful method to simultaneously constrain the
stellar mass completeness and the SHMR for current and future
surveys. The constraining power of the method comes from the accurate
measurements of the observed SMFs and galaxy 2PCFs, which rely on the
large sample volumes of the galaxy surveys. As shown in the previous
section and discussed in G18, the final model constraints are
relatively independent of the detailed functional forms of the ICSMF
model ingredients, once they are flexible enough to account for the
selection effects in the real galaxy survey data.

Different from the many previous deep galaxy surveys with photometric
redshifts, the eBOSS ELG survey has obtained accurate redshift
information for most of the observed galaxies. By modeling the ELG
samples at different redshift intervals independently, we can reliably
study the evolution of the star-forming galaxies without assuming any
redshift dependency in the model parameters, which may distort the
high-redshift results in order to fit the low-redshift
measurements. However, limited by the low completeness of the eBOSS ELG
sample, the clustering measurements at these redshifts are still
noisy. The resulting constraints on the SHMRs and intrinsic SMFs are
not as accurate as those at low redshifts with volume-limited samples
\citep{Zehavi11,Yang12}. But the ICSMF model still provides reasonable
constraints to the galaxy populations at these redshifts.

We note that in principle, the completeness of galaxy populations
would depend on their stellar mass, color, and other physical
properties. What we obtain in the ICSMF model should be regarded as an
average sample completeness at a given stellar mass. Our current eBOSS
data are not accurate enough to fully break the degeneracies. We
expect the ICSMF model to provide tighter constraints in future
large-scale galaxy surveys.

\section{Conclusions} \label{sec:conclusion}

In this paper, we apply the ICSMF model introduced by G18 to the eBOSS
ELG samples over the redshift range of $0.7<z<1.2$ for galaxies with
$10^{10}\msun<M_*<10^{11.6}\msun$. By fitting to the observed galaxy
clustering and SMF measurements, we are able to constrain the sample
completeness, the SHMR and the quenched galaxy fraction at the same
time, which serves as a powerful way to study galaxy evolution using
the large-scale galaxy surveys. The intrinsic galaxy SMFs are then
directly inferred from the ICSMF model for each galaxy sample at
different redshifts.

Our main conclusions are summarized as follows.

\begin{itemize}
\item The average stellar mass completeness of the eBOSS ELG samples
  varies from 1\% to 10\% for different redshift samples. The galaxy
  samples at $z<1$ are slightly more complete, compared to the higher
  redshift ones.
	
\item There is only weak evolution of the SHMR for ELGs in the
  redshift range of $0.7<z<1.2$ for low-mass halos of $M<10^{13}\msun$, while the high-mass end slope $\alpha$
  becomes flat for $z>0.8$.
	
\item We have obtained the intrinsic SMFs for ELGs in the four
  redshift bins in the range of $0.7<z<1.2$, and the SMF for total
  galaxies in the redshift bin of $0.7<z<0.8$.

\item The low-mass end ($M_*<10^{11}\msun$) of the galaxy SMFs for the
  star-forming galaxies is roughly unchanged from $z=1.2$ to
  $z=0.7$. 
	
\item The peak halo mass distribution of the central ELGs is around
  $M\sim10^{12}\msun$, while that of the satellite ELGs increases to
  $M\sim10^{12.6}\msun$.
	
\item The satellite fraction of the eBOSS ELG varies from $13\%$ to
  $17\%$ at different redshifts and the average galaxy bias increases
  from $1.1$ at $z\sim0.7$ to $1.4$ at $z\sim1.2$.
	
\end{itemize}

\section*{Acknowledgements}

This work is supported by the National Key R\&D Program of China
(grant Nos. 2015CB857003, 2015CB857002), national science foundation
of China (Nos. 11621303, 11655002, 11773049, 11833005, 11828302).
H.G. acknowledges the support of the 100 Talents Program of the
Chinese Academy of Sciences. This work is also supported by a grant
from Science and Technology Commission of Shanghai Municipality
(Grants No. 16DZ2260200).

We thank the anonymous reviewer for the helpful comments that significantly improved the presentation of this paper.
We thank Yen-Ting Lin for carefully reading the manuscript and
providing detailed comments. We thank Rita Tojeiro for useful
discussions.  We gratefully acknowledge the use of the High
Performance Computing Resource in the Core Facility for Advanced
Research Computing at the Shanghai Astronomical Observatory. We
acknowledge the Gauss Centre for Supercomputing
e.V. (www.gauss-centre.eu) and the Partnership for Advanced
Supercomputing in Europe (PRACE; www.prace-ri.eu) for funding the
MultiDark simulation project by providing computing time on the GCS
Supercomputer SuperMUC at Leibniz Supercomputing Centre (LRZ,
www.lrz.de).

Funding for the Sloan Digital Sky Survey IV has been provided by the
Alfred P. Sloan Foundation, the U.S. Department of Energy Office of
Science, and the Participating Institutions. SDSS acknowledges support
and resources from the Center for High-Performance Computing at the
University of Utah. The SDSS web site is www.sdss.org.

SDSS is managed by the Astrophysical Research Consortium for the
Participating Institutions of the SDSS Collaboration including the
Brazilian Participation Group, the Carnegie Institution for Science,
Carnegie Mellon University, the Chilean Participation Group, the
French Participation Group, Harvard-Smithsonian Center for
Astrophysics, Instituto de Astrofísica de Canarias, The Johns Hopkins
University, Kavli Institute for the Physics and Mathematics of the
Universe (IPMU) / University of Tokyo, Lawrence Berkeley National
Laboratory, Leibniz Institut für Astrophysik Potsdam (AIP),
Max-Planck-Institut für Astronomie (MPIA Heidelberg),
Max-Planck-Institut für Astrophysik (MPA Garching),
Max-Planck-Institut für Extraterrestrische Physik (MPE), National
Astronomical Observatories of China, New Mexico State University, New
York University, University of Notre Dame, Observatório Nacional /
MCTI, The Ohio State University, Pennsylvania State University,
Shanghai Astronomical Observatory, United Kingdom Participation Group,
Universidad Nacional Autónoma de México, University of Arizona,
University of Colorado Boulder, University of Oxford, University of
Portsmouth, University of Utah, University of Virginia, University of
Washington, University of Wisconsin, Vanderbilt University, and Yale
University.

\appendix

\section{Robustness of Model Predictions}\label{subsec:test}

Compared to G18, we introduce an additional quenched halo fraction
$f_{\rm q}(M)$ to model the ELGs in eBOSS. Since we have assumed a
simple functional form of Eq.~\ref{eq:fsf} to model the variation of
quenched fraction with the halo mass, it is worth checking the effect
of different $f_{\rm q}(M)$ models. For comparison, we choose another
$f_{\rm q}(M)$ model proposed by \cite{Zu16} as follows:
\begin{equation}
f_{\rm q}(M)=1-\exp[-(M/M_{\rm q})^\mu]\label{eq:fq2},\\
\end{equation}
where the two free parameters are $M_{\rm q}$ and $\mu$. In the
following, we refer to this model as ``$f_{\rm q}(M)$ model 2'', while
our fiducial model is referred to as ``$f_{\rm q}(M)$ model 1''.

Another potential source of uncertainty is that we have assumed the
same scatter $\sigma_*$ in Eq.~\ref{eq:csmf} for both the quiescent
and star-forming galaxies. \cite{Tinker13} had modeled the scatter
$\sigma_*$ for the quiescent and star-forming galaxies independently
using the measurements of the angular clustering and galaxy-galaxy
lensing. They found a $\sigma_*$ value of $0.18\pm0.05$ for quiescent
galaxies and $0.25\pm0.01$ for star-forming galaxies at
$0.74<z<1$. Although we have a comparable scatter for the quiescent
galaxies, our adopted scatter for the star-forming galaxies
($\sigma_*=0.173$ at $0.7<z<0.8$) is significantly smaller. In order
to test the effect of the assumed scatter, we further include a model
similar to our fiducial one, but we allow $\sigma_*$ for the
star-forming galaxies to be a free parameter, which is referred to as
``$f_{\rm q}(M)$ model 1+scatter''. In this model, we still assume a
constant scatter of $\sigma_*=0.173$ for the quiescent galaxies.

\begin{figure}
	\centering
	\includegraphics[width=0.47\textwidth]{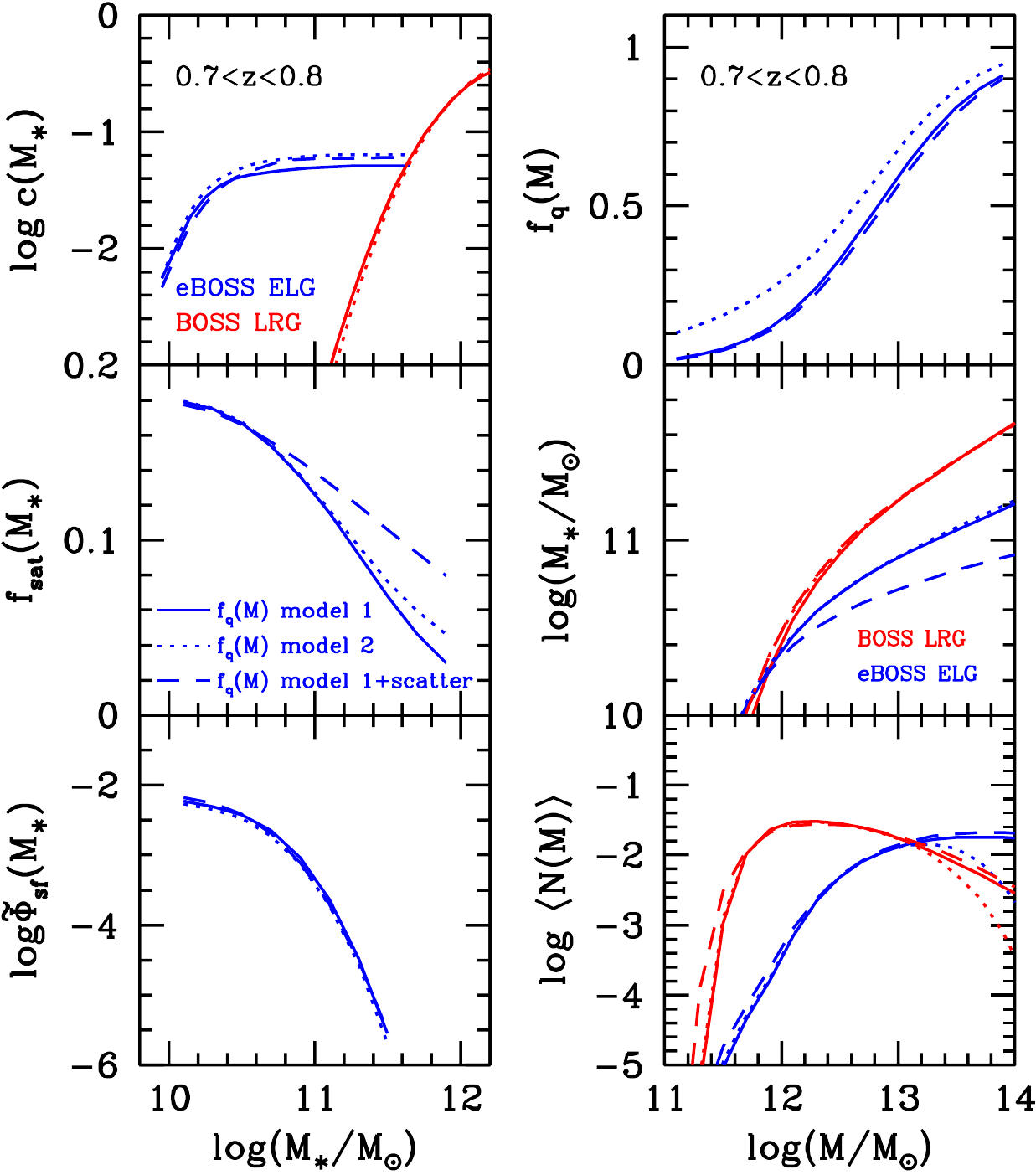}
	\caption{Comparisons of model predictions for our fiducial
          model (``$f_{\rm q}(M)$ model 1'') and two variations of
          ``$f_{\rm q}(M)$ model 2'' and ``$f_{\rm q}(M)$ model
          1+scatter'' (see text for details). The comparisons are made
          for the stellar mass completeness $c(M_*)$, quenched
          fraction $f_{\rm q}(M)$, satellite fraction
          $f_{\rm sat}(M_*)$, SHMR, intrinsic SMF
          $\tilde{\Phi}_{\rm sf}(M_*)$, and the halo occupation
          functions $\langle N(M)\rangle$. The solid, dotted and
          dashed lines are for the ``$f_{\rm q}(M)$ model 1'',
          ``$f_{\rm q}(M)$ model 2'' and ``$f_{\rm q}(M)$ model
          1+scatter'', respectively.}
	\label{fig:modelcomp}
\end{figure} 

We show in Figure~\ref{fig:modelcomp} the comparisons of best fits
from the above three models using the BOSS LRG and eBOSS ELG
measurements at $0.7<z<0.8$. The comparisons are made for the stellar
mass completeness $c(M_*)$, quenched fraction $f_{\rm q}(M)$,
satellite fraction $f_{\rm sat}(M_*)$, SHMR, intrinsic galaxy SMF of
$\tilde{\Phi}_{\rm sf}(M_*)$, and the halo occupation functions
as in previous figures. The solid, dotted and
dashed lines are for the ``$f_{\rm q}(M)$ model 1'', ``$f_{\rm q}(M)$
model 2'' and ``$f_{\rm q}(M)$ model 1+scatter'', respectively.

The best-fitting $\sigma_*$ for the ``$f_{\rm q}(M)$ model 1+scatter''
is $\sigma_*=0.22^{+0.05}_{-0.04}$, basically consistent with our
model assumption within the errors. The scatter is not tightly
constrained with the current eBOSS ELG sample due to the noisy
clustering measurements. The inclusion of $\sigma_*$ as a free
parameter makes the high-mass end slope of the SHMR
shallower, which increases the satellite fraction at the massive end,
as galaxies at a given stellar mass would have higher chances to
populate the massive halos. However, other model predictions stay
roughly the same with a larger $\sigma_*$.

Adopting the ``$f_{\rm q}(M)$ model 2'' would affect the quenched
fraction at the $\sim 10\%$ level, with more halos hosting quenched
central galaxies, especially for low-mass halos. However, as the constrained stellar mass completeness is relatively higher in order to fit the observed SMF of $\Phi_{\rm sf}(M_*)$, the final intrinsic SMF
$\tilde{\Phi}_{\rm sf}(M_*)$ of the different models still have very
good agreement with each other. As discussed previously, the exact
high-mass end slopes of the halo occupation functions for both central
and satellite galaxies would depend on the detailed model choice of
$f_{\rm q}(M)$, while the differences at the low-mass end are relatively
small. As the most massive halos only have minor contributions to the
overall halo mass function, the average host halo mass distributions
are consistent among different models. The SHMR of ``$f_{\rm q}(M)$
model 2'' is still in reasonable agreement with our fiducial model.

In addition, although we have applied a constant $M_{\rm q}$ value constrained from $0.7<z<0.8$ to higher redshift bins, our model constraints at these redshifts stay roughly the same for varying $M_{\rm q}$ values within the 1$\sigma$ range of $12.83\pm0.24$, similar to the situation with changing the $f_{\rm q}(M)$ models. While the SHMR can be constrained by the clustering measurements, changing the $f_{\rm q}(M)$ will be compensated by the corresponding changes in the stellar mass completeness functions as shown in Eqs.~(\ref{eq:ncen_int})--(\ref{eq:phisat}).


\end{document}